\documentclass[a4paper, 11pt]{article}

\usepackage{graphicx}
\usepackage{amsmath, amssymb, bm, longtable, array, tabularx, subcaption, braket, listings,siunitx,appendix,verbatim,cite,xcolor}
\usepackage{seqsplit}
\allowdisplaybreaks
\usepackage{hyperref}
\usepackage{cleveref}

\newcommand{\C}{\mathbb{C}}
\newcommand{\IC}{\mathbb{C}}
\newcommand{\IZ}{\mathbb{Z}}

\newcommand{\II}{\mathbb{I}}
\newcommand{\ep}[1]{\emph{#1}}
\newcommand{\be}{\begin{equation}}
\newcommand{\ee}{\end{equation}}
\newcommand{\nn}{\nonumber}
\newcommand{\gen}[1]{\langle #1 \rangle}
\newcommand{\cM}{{\cal M}}
\newcommand{\cX}{{\cal X}}
\newcommand{\cG}{{\cal G}}
\newcommand{\cR}{{\cal R}}

\renewenvironment{thebibliography}[1]{%
\begin{oldthebibliography}{#1}%
\setlength{\parskip}{0ex}%
\setlength{\itemsep}{0ex}%
}%
{%
\end{oldthebibliography}%
}

\newtheorem{theorem}{Theorem}[section]
\newtheorem{definition}{\bf DEFINITION}[section]
\newtheorem{proposition}{\bf PROPOSITION}[section]

\newcommand{\setall}{\setcounter{equation}{0}
        \setcounter{theorem}{0}}

\renewcommand{\thefootnote}{\fnsymbol{footnote}}

\begin{document}
\thispagestyle{empty}
\renewcommand{\thefootnote}{\fnsymbol{footnote}}

\begin{titlepage}

\vspace{0.2in}

\begin{center}

\textbf{
{\LARGE Standard Model Plethystics}
}

\vspace{0.2in}

\textbf{
Yan Xiao$^{a}$\footnote{\texttt{Yan.Xiao@city.ac.uk}},
Yang-Hui He$^{a,b,c}$\footnote{\texttt{hey@maths.ox.ac.uk}},
Cyril Matti$^{a}$ \footnote{\texttt{cyril.matti@a3.epfl.ch}}\\
}

\vspace{0.1in}

${}^a$ \textit{Department of Mathematics, City, University of London,\\
Northampton Square, London EC1V 0HB, UK}
\vskip0.25cm

${}^b$ \textit{Merton College, University of Oxford, OX1 4JD, UK}
\vskip0.25cm

${}^c$ \textit{School of Physics, NanKai University, Tianjin, 300071, P.R.\ China,}
\vskip0.25cm

\end{center}

\vspace{10mm}

\begin{abstract}
We study the vacuum geometry prescribed by the gauge invariant operators of the MSSM via the Plethystic Programme.  This is achieved by using several tricks to perform the highly computationally challenging Molien-Weyl integral, from which we extract the Hilbert series, encoding the invariants of the geometry at all degrees. The fully refined Hilbert series is presented as the explicit sum of 1422 rational functions. We found a good choice of weights to unrefine the Hilbert series into a rational function of a single variable, from which we can read off the dimension and the degree of the vacuum moduli space of the MSSM gauge invariants. All data in {\sc Mathematica} format are also presented.
\end{abstract}

\end{titlepage}

\tableofcontents

\vspace{1in}

\section{Introduction and Summary}\label{sec:intro}
The Standard Model of particle theory containing specific gauge interactions is expected to have more structures when
extended to energies above 1-10 TeV, where supersymmetry might be incorporated. 
Indeed, to derive the Standard Model as an effective theory from a unified theory containing gravity is one of the chief prospects of theoretical particle physics.
One of the most important aspects of a supersymmetric gauge theory is that its vacuum, due to the omnipresence of scalars in the theory, can be highly non-trivial, as parametrized by the vacuum expectation values (VEVs) of gauge invariant operators (GIOs) composed of these scalar fields \cite{Buccella:1982nx,Gatto:1986bt,Procesi:hr}.
This vacuum moduli space (VMS) can be explicitly obtained as solution of constraints coming from F-flatness and D-flatness and be realized, in the language of algebraic geometry, as an {\it algebraic variety} \cite{lt}.

Supersymmetric extensions of the Standard Model clearly constitute one of the central subjects in particle phenomenology.
In particular, the minimal extension, the MSSM, and its variants, have been subject to intense investigations.
The flat directions of the MSSM have been identified in \cite{Gherghetta:1995dv}.
Combining these directions of thought, a long programme was launched to study the vacuum geometry of the MSSM and its relatives
\cite{Gray:2005sr,Gray:2006jb,Gray:2008yu,He:2014loa,He:2014oha,He:2015rzg}: under the guiding principle that ``interesting geometry is coextensive with interesting physics'', the ultimate goal is to use geometric and topological properties of VMS as a selection rule for operators in the Standard Model Lagrangian. 
Specifically, if the VMS were to be found to have some special form in the mathematical sense, which (1) cannot be
explained in terms of symmetries relating the relevant degree of freedom in the low energy effective field theory; and (2)
is very unlikely to have occurred by chance, then this special form should be regarded as a consequence of some
unknown physics. In this setting, we take \ep{special} to mean non-trivial properties of algebraic geometry, such as exhibited by  interesting topological invariants or emergence of special holonomy.

Under such a spirit, the presence of any
special geometry would be a collective consequence of factors such as gauge group, particle spectrum and the
interactions within the theory. Therefore, if a special geometry is found within the low energy sector of a theory and
this geometry is very unlikely to have arisen by chance, then the existence of such geometry in the VMS should be a
fundamental property across all energy scales. 
Hence, the addition of higher dimensional operators to our theory
can only occur when they are compatible with this structure. In such sense, we are placing very restrictive constraints
on allowed physical processes that are mediated by certain operators.

Already, many interesting features have been found, such as the VMS of the electro-weak sector being an affine cone over the classical Veronese surface, a structure ruined by addition of R-parity-violating operators, or the sensitive dependence of the geometry on the number of generations, or the appearance of Calabi-Yau varieties, etc \cite{He:2014loa,He:2014oha,He:2015rzg}.
Supersymmetry and the VMS thus provide us with a low energy window of how geometry can guide certain phenomenological questions. 
A good analogy would be the study of complex numbers: many unforeseen and crucial properties of analytic functions are visible only after the complex extension of the real numbers, so too would important properties of quantum field theories - and the Standard Model in particular - only be visible after the supersymmetric extension. In this light, regardless of whether there exist supersymmetric particles, the study of supersymmetric structures and the VMS is an integral part of the study of field theory.

Despite the progress made over the years in the aforementioned programme, an important question has remained: ``what is the VMS of the MSSM?''
The reason is purely computational: in component form, there are about 50 scalars and about 1000 GIOs, which by the state-of-art standard procedure  in algebraic geometry adapted to calculate the VSM \cite{He:2014oha}, is beyond the computational power of even the most sophisticated computers by direct means.
Therefore, investigations thus far have focused on the electro-weak sector wherein, as discussed, so much have already been uncovered.
It is indeed expected that the geometry of the full MSSM would have far richer and salient features.

Rather fortuitously, there has been a parallel programme in studying supersymmetric gauge theories: this is the so-called 
{\it Plethystic Programme} \cite{Benvenuti:2006qr,Feng:2007ur}.
It originated in the study of quiver gauge theories which arise from string theory, as world-volume theories on D-branes probing Calabi-Yau singularities \cite{Feng:2000mi}, which have become the playground for the AdS/CFT correspondence \cite{Maldacena:1997re}.
Here, the VMS of the gauge theory is, by construction, the affine Calabi-Yau variety transverse to the D-branes, and was in part the initial motivation for the VMS/phenomenology programme.
It allows one to build criteria to rule out certain top-down string model building to obtain desired low-energy outcomes: if the VMS is \ep{not} 
Calabi-Yau, then one cannot use a direct top-down method.

The central object to the Plethystic Programme is the Hilbert series, well-known to algebraic geometry as the generating function for counting the dimension of graded pieces of the coordinate ring. 
Harnessing this analogy with the super-conformal index \cite{Pouliot:1998yv,Romelsberger:2005eg,Romelsberger:2007ec,Dolan:2007rq,Dolan:2008qi,Hanany:2006uc,Sundborg:1999ue,Brown:2007xh}, the original motivation was to study the chiral ring of BPS operators in supersymmetric gauge theories: the Hilbert series the counts the single-trace operators, whilst its plethystic exponential counts the multi-trace, and its plethystic logarithm encodes the generators and relations of the variety.
A host of activity ensued, developing and refining various aspects of the programme \cite{Hanany:2008sb,Hanany:2008kn,Forcella:2007wk,Hanany:2007ap,Balasubramanian:2007hu,Butti:2007jv,Forcella:2008bb,Cremonesi:2017jrk,Cremonesi:2016nbo,Hanany:2014hia,Dey:2013fea,Rodriguez-Gomez:2013dpa,Jokela:2011vg}, even using the Hilbert-series technology to (the regular, non-supersymmetric) Standard-Model phenomenology \cite{Lehman:2015via,Hanany:2010vu,Lehman:2015coa,Henning:2015daa}.

A natural question therefore arises as to whether the two programmes can come to a useful syzygy.
Specifically, can certain properties of VMS for the MSSM be obtained without recourse to the computationally expensive elimination algorithm, but be deduced from the Hilbert series, which may be calculated via other means?
Luckily, this is indeed the case.
When the algebraic variety has extra symmetries, such as precisely in our cases, when they come from certain symplectic quotients of Lie-group invariants, there is a classic method of Molien-Weyl integration \cite{Fulton:1991rt} to obtain the Hilbert series.
The purpose of this paper is to perform this, albeit difficult, integral and obtain the Hilbert series explicitly for the MSSM, whence one can further deduce relevant geometrical quantities.

The objects that immediately follow from the Hilbet series are the degree and dimension of the vacuum moduli space. In this work, we obtain the dimension of the VMS to be 40 and the degree given by the product of the following prime factors
\begin{equation*}\label{eq:deg}
\begin{split}
&2^{898} \cdot 3^{324}\cdot 5^{145}\cdot 7^{120}\cdot 11^{58}\cdot 13^{53}\cdot 17^{31}\cdot 19^{35}\cdot 23^{21}\cdot 29^{16}\cdot 31^{14}\cdot 37^{17}\cdot 41^{10}\cdot 43^{11}\cdot 47^{13}\cdot 53^7\cdot 59^8\cdot\\
&61^{10}\cdot
67^6\cdot 71^4\cdot 73^{11}\cdot 79^6\cdot 83^6\cdot 89^5\cdot 97^5\cdot 101^2\cdot 103^4\cdot 107^3\cdot 109^3\cdot 113^2\cdot 127^3\cdot 131^3\cdot 137^2\cdot 139\cdot\\
&149^3\cdot
151^3\cdot 157^2\cdot 163^5\cdot 167\cdot 173\cdot 179^3\cdot 181^3\cdot 191\cdot 193^5\cdot 197^2\cdot 199\cdot 211^2\cdot 229\cdot 251\cdot 257^2\cdot 263\cdot\\
&269\cdot 271^3\cdot 277\cdot 283^2\cdot 311\cdot 313\cdot 331\cdot 337\cdot 353\cdot 373^3\cdot 379\cdot 389\cdot 431\cdot 433\cdot 443\cdot 461\cdot 467\cdot 491\cdot\\
&509\cdot 521\cdot 541^2\cdot 547^2\cdot 557^3\cdot 563^2\cdot 587\cdot 599\cdot 607\cdot 643^3\cdot 727\cdot 757\cdot 773^3\cdot 811\cdot 821\cdot 977\cdot 1061\cdot\\
&1151\cdot 1279\cdot 1531\cdot 1549\cdot 1571\cdot 1579~,
\end{split}
\end{equation*}
with the following number that does not contain any prime factors between $1579$ and number of order $10^{11}$,\footnote{This number itself is not necessarily a prime but factorisation of this number beyond primes of order $10^{11}$ is out of the capabilities of normal laptop.}
\newline
\seqsplit{11196186329560947241455148908824054684468743740728908934170824241971190830192362923499173415152896923453222633515598618927623339056067805499505607184567519814957213507242680912973888648108680864649889603020633846308350639535586019348337761614905562118460688361209927517001930882524733517050667597849316746015467332076233464803805765046927548756022781332812982563552464841943947455364952170146966153056751240559527427052868262882098801432408663132559876247420465532292619426459726799545486344638850079630506164699}.

The organization of the paper is as follows.  Section \ref{sec:PP} reviews some elements of the {\it Plethystic Programme} and we can see therein how the it establishes the connection between Hilbert series and
the geometry of VMS. Section \ref{sec:example} gives examples illustrating the programme both in sQCD and Abelian gauge theory. Section \ref{sec:setup} establishes the scene for the plethystic
integral for MSSM using the related characters of $SU(3)$, $SU(2)$ and $U(1)$ as well as corresponding charges for the matter content thereof. Section \ref{sec:MSSM_HS} gives the description
of obtaining the Hilbert series for MSSM with certain subtleties and the main obstacles within this procedure. The results are also presented with more details in this section as well.

Lastly, the VMS obtained here is not constrained by the superpotential $W$ of MSSM, {\it i.e.} the relations from requiring $\partial W/\partial \phi_i = 0$ with 
$\phi_i$ being the scalar component of the chiral fields in MSSM, are not imposed on reaching the VMS. The case of non-trivial superpotential $W \neq 0$
is therefore left for future work.
\section{The Plethystic Programme}\label{sec:PP}\setall
In this section, we review some aspects of the Plethystic Programme.
The reader is also referred to \cite{He:2016fnb} for a rapid review of the programme and its context within quiver representations and gauge theory.

\subsection{Elimination Algorithm for VMS}\label{sec:M}
We first briefly recall the algorithm for computing the VMS of a generic $\mathcal{N}=1$ supersymmetric gauge theory with gauge group $\cG$, fields whose scalar components are $\phi_i$, and a polynomial superpotential $W$ therein.
The most efficient method to obtain the VMS is as follows.
\begin{itemize}
\item {\bf INPUT:}
\begin{enumerate}
\item Superpotential $W(\{\phi_i\})$, a polynomial in variables $\phi_{i=1, \ldots, n}$.
\item Generators of GIOs: $r_j(\{\phi_i\})$, $j=1, \ldots, k$ polynomials in $\phi_i$ invariant under $\cG$.
\end{enumerate}

\item {\bf ALGORITHM:}
\begin{enumerate}
\item Define the polynomial ring $R = \IC[\phi_{i=1,\ldots,n}, y_{j=1,\ldots,k}]$.
\item Consider the ideal $I = \gen{\frac{\partial {W}}{\partial \phi_i}, y_j - r_j(\{\phi_i\})}$.
\item Eliminate all variables $\phi_i$ from $I \subset R$, giving the ideal $\cM$ in terms of $y_j$.
\end{enumerate}

\item {\bf OUTPUT:}\\
$\cM$ corresponds to the vacuum moduli space as an affine variety in $\IC[y_1, \ldots, y_k]$.
\end{itemize}
In the ensuing, we will address general varieties $\cX$ though ultimately we will specialize to when $\cX$ is obtained as the VMS $\cM$ from the above.

\subsection{The Hilbert Series}
Now we define the protagonist of our investigations.
\begin{definition}
Given an algebraic variety $\mathcal{X}$ in $\C[x_1,\dots,x_n]$, the Hilbert series is the generating function
\begin{equation}
H_{\cX}(t) = \sum_{i=0}^\infty ({\rm dim}_\C \mathcal{X}_i)t^i \quad,
\end{equation}
where $\mathcal{X}_i$ the $i$-th graded piece of the coordinate ring for $\mathcal{X}$ and can be regarded as the number of independent degree $i$
(Laurent) polynomials on $\mathcal{X}$. 
\end{definition}
Note that the Hilbert series is not a topological invariant and it depends on the embedding of $\mathcal{X}$.
Of course, $H(t)$ can be generalized to be multi-variate $H(t_1,\ldots,t_n)$ by considering the multi-graded pieces 
$\mathcal{X}_{i_1, \ldots, i_n}$.
The dummy variables $t_i$ are called {\bf fugacities} in the physics literature.
When there is more than one variable $t_i$, the Hilbert series is called {\it refined}, otherwise it is often called {\it unrefined}.

There are two important forms of Hilbert series which will be used later in this paper for obtaining the 
degree and dimension of the underlying variety. 
We have that (cf.~\cite{Cox:1992ca,Schenck:2003ca})
\begin{theorem}\label{thm:HS-dim-deg}
The Hilbert series $H(t)$ is a rational function in $t$ and can be written in two ways:
\begin{equation}\label{def:HS}
H_{\cX}(t) = 
\begin{cases}
\frac{Q(t)}{(1-t)^k} \quad,& \text{Hilbert series of the first kind\quad;}\\
\frac{P(t)}{(1-t)^{\text{dim}(\mathcal{X})}} \quad, &\text{Hilbert series of the second kind\quad.}
\end{cases}
\end{equation}
Here both $P(t)$ and $Q(t)$ are polynomials with \emph{integer} coefficients and the dimension of the embedding 
space is given by the power of the denominators. 
Moreover, $P(1) = {\rm degree}(\mathcal{X})$.
\end{theorem}
Thus we have a convenient way to obtain the degree of the variety \footnote{
Recall that when an ideal is a single polynomial, i.e., $\mathcal{X}$ is a hypersurface, the degree of the
variety is simply the degree of the polynomial. For multiple polynomials, the degree is a generalisation of this notion.
It then becomes the number of intersection points between a generic line and the variety. 
}.

Furthermore, since the Hilbert series is a rational function,  
\begin{theorem}\label{thm:HS-partial-deg}
$H(t)$ affords a partial-fraction expansion around $t=1$ \cite{FOS,Martelli:2006yb}
\begin{equation}
H_{\cX}(t) = \frac{P(1)}{(1-t)^{{\rm dim}(\mathcal{X})}} + \dots \ .
\end{equation}
Thus, the coefficient of the leading pole gives the degree of the variety while the order of the pole is the dimension.
\end{theorem}
Indeed, for Calabi-Yau varieties, the coefficient of the leading pole can also be interpreted as the volume of base Sasaki-Einstein manifold,
which in the AdS/CFT context is related to the central charges of the supersymmetric gauge theory \cite{Martelli:2006yb,Forcella:2008bb,Gauntlett:2006vf,He:2017gam}.

\paragraph{Remark: }
We remark that when $\mathcal{X}$ is a quotient variety, i.e., $\mathcal{X} \simeq \IC^n / \Gamma$ for some discrete finite group $\Gamma$ acting on the $n$ coordinates of $\IC^n$, the problem of computing $H(t; \IC^n / \Gamma)$ reduces to counting the number of algebraically independent polynomials of each degree that are invariant under group action. 
This problem was solved by Molien \cite{molien,Cox:1992ca} and the corresponding Hilbert series is the well-known {\bf Molien series}, which can be computed by a sum over group elements:
\begin{equation}\label{molienseries}
H_{\cX}(t) = \frac{1}{|\Gamma|} \sum\limits_{g \in \Gamma} \det(\II - t g)^{-1} \ .
\end{equation}

\subsection{Molien-Weyl Formula}\label{s:MW}
The case of our principle interest is when $\cX$ is not a finite quotient, but of the form of a symplectic quotient by a (continuous) Lie group coming from gauge symmetry.
Luckily, there is a generalization of \eqref{molienseries} into a so-called {\bf Molien-Weyl Integral} \cite{Fulton:1991rt} (cf.~\cite{Gray:2008yu,Hanany:2008kn,Hanany:2008sb}).
The problem of finding invariants under continuous gauge group is at the heart of invariant theory that can be traced back to 19-th century and we present a rapid review of the origin of Hilbert series and Gr\"obner basis in the context of commutative algebra in \cref{append:com_alg}.

For our incarnation in physics, we wish to compute the Hilbert series for $\cX = \cM$ coming from the algorithm in \S\ref{sec:M}, whose coordinate ring $\mathcal{R}$ is the projection of the quotient ring $R/I$ onto $\IC[y_j]$.
Now, the complexified gauge group $\mathcal{G}_c$ and global symmetry group act naturally on $\mathcal{R}$ and we can grade the elements therein with gauge and global charges. 
Let us denote the global Abelian charge as $t_i$ and that of Cartan subgroup of the gauge group $\mathcal{G}_v$ by $z_i$. 
This gives the generating function, i.e., the Hilbert Series, of the graded ring $\mathcal{R}$ as
\begin{equation}
H_{\mathcal{R}}(t; z) = \sum_{n, m} a_{nm}z^nt^m \quad,
\end{equation}
which can be written as a power series in the global charge $t$ and a Laurent expansion in the gauge charges $z$. 

Since the gauge symmetry commutes with global symmetry, all elements of $\mathcal{R}$ with given charge $t^m$ should form a representation
$\chi^m$ of $\mathcal{G}_c$:
\begin{equation}
H_{\mathcal{R}}(t,z) = \sum_{m=0}^\infty \chi^m(z)t^m = \sum_{m=0}^\infty \left(\sum_i a_i^m\chi^{(i)}(z)\right)t^m \quad.
\end{equation}
In the last step, we have decomposed $\chi^m$, the representation on the elements of charge $t^m$, into irreducible representations
$\chi^{(i)}$. 
Therefore, the generating function for invariants is given by the projection onto the trivial representation with character
$\chi^{(0)} = 1$,
\begin{equation}
H_{\mathcal{R}}^{inv}(t) = \sum_{m=0}^\infty a_0^mt^m \quad.
\end{equation}
The projection is done by averaging $H(t;z)$ on the gauge group with Haar measure $d\mu(z)$,
\begin{equation}
\int d\mu(z) \chi^{(i)}(z) = \delta_{i,0} \quad.
\end{equation}
Explicitly, a group $G$ of rank $r$ has its Haar measure in terms of contour integral
\begin{equation}
\frac{1}{|W|}\prod_{i=1}^r \int_{|z_i| = 1} \frac{dz_i}{2\pi i z_i} \prod_{\alpha \in \Delta}(1 - \prod_i^r z_i^{\alpha_i}) ~,
\end{equation}
where $|W|$ is the order of the Weyl group and $\alpha$ is a root, or the weight of the adjoint
representation such that $\alpha_i$ is the  $i$-th entry of the weight vector in the Dynkin basis.

Putting all the above together, the Molien-Weyl formula for the Hilbert series of the variety $\cM$ whose coordinate ring is $\cR$ reads
\begin{equation}\label{def:molien}
H_{\mathcal{R}}^{inv}(t) =\frac{1}{|W|}\prod_{i=1}^r \int_{|z_i| = 1} \frac{dz_i}{2\pi i z_i} \prod_{\alpha \in \Delta}(1 - \prod_i^r z_i^{\alpha_i}) H_{\mathcal{R}}(t;z) 
\quad .
\end{equation}
Note here the integration requires knowledge about the Hilbert series of the coordinate ring $\cR$.

\subsection{Plethystics and Syzygies}
The next crucial concept needed is that of \ep{Plethystics}. 
\begin{definition}
Let $g(t_1,\dots,t_n)$ be a multivariate analytic function.
The \emph{Plethystic Exponential} is
\begin{equation}\label{def:PE}
\text{{\rm  PE}}[g(t_1,\dots,t_n)] := \exp \left( \sum_{k=1}^\infty \frac{g(t_1^k,\dots,t_n^k) - g(0,\dots,0)}{k} \right)
\quad.
\end{equation}
\end{definition}
It is easy to show (q.v.,~\cite{Feng:2007ur}) that (being an exponential) the plethystic exponential is multiplicative in additive arguments, and furthermore
\begin{equation}
f(t) = \sum\limits_{n=0}^\infty a_n t^n \quad \Rightarrow \quad
PE[f(t)] = \exp\left( \sum_{n=1}^\infty \frac{f(t^n) - f(0)}{n} \right) = \prod\limits_{n=1}^\infty (1-t^n)^{-a_n} \ .
\end{equation}
The product form is particularly useful and it is usually called {\bf Euler form}.

It is a non-trivial fact \cite{Fulton:1991rt,Feng:2007ur} that this has an analytic inverse function called the plethystic logarithm
\begin{equation}
\text{PE}^{-1}[g(t_1,\dots,t_n)] =  \sum_{k=1}^\infty \frac{\mu(k)}{k} \log (g(t_1^k, \ldots, t_n^k)) \ , 
\end{equation}
where 
\begin{equation}
\mu(k) := \left\{\begin{array}{lcl}
0 & & k \mbox{ has repeated prime factors}\\
1 & & k = 1\\
(-1)^n & & k \mbox{ is a product of $n$ distinct primes} 
\end{array}\right. 
\end{equation}
is the M\"obius mu-function.

Remarkably, the plethystic logarithm can be used to find the defining relation (syzygies) of  the generators of an algebraic variety \cite{Benvenuti:2006qr,Feng:2007ur}. 
\begin{proposition}\label{plog}
  Given Hilbert series $H(t; \cX)$ of an algebraic variety $\cX$, the plethystic logarithm is of the form
  \[
  {\rm PE}^{-1}[H(t; \cM)] = b_1 t + b_2 t^2 + b_3 t^3 + \ldots
  \]
  where all $b_n \in \IZ$ and a positive $b_n$ corresponds to a generator in coordinate ring of $\cX$ and a negative $b_n$, a relation.
  In particular, if $\cX$ is complete intersection, then $PE^{-1}[H(t; \cM)]$ is a finite polynomial.
\end{proposition}
We illustrative this proposition in detail with concrete examples in Appendix \ref{ap:plog}.

\subsection{Summary}

To summarise, we recall that the plethystic exponential (PE) is defined to be
\begin{equation}\label{def:PExp}
{\rm PE} \left[ \chi^G_R (z_a) \sum_i^{N_f} t_i \right] \equiv {\rm exp} \left[ \sum_{k=1}^\infty  \sum_{i=1}^{N_f}
\frac{1}{k} \left( t_i^k \chi^G_R (z_a^k)\right) \right] ~,
\end{equation}
where $\chi_R^G(t_i, z_a)$ is the character for representation $R$ of group $G$ and it is expanded into monomials of complex variables
$z_i$. Note that the number of complex variables $z_i$ is equal to the rank of group $G$. The expansion of PE gives the complete set
of combinations of ``fugacities" $t_i$. To find the generating function of gauge invariant opertors under group $G$, we need to project
the representation generated by PE onto trivial subrepresentations of $G$.  This is can be carried out by integrating over the whole group.
This is precisely the Hilbert series $H^{inv}_{\mathcal{R}}$ in \cref{def:molien} whose {\bf Molien-Weyl} formula for Plethystic Integral is given by
\begin{equation}
g = \int_G {\rm d}\mu_G ~ {\rm PE} \left[ \chi^G_R (z_a) \sum_i^{N_f} t_i \right] ~,
\end{equation}
where ${\rm d}\mu_G$ is the Haar measure for group $G$. With these data at hand, we conveniently package them into the following theorem

\begin{theorem} [Molien-Weyl Integral for the Hilbert Series] \label{MW}
Given gauge group $G$, with Haar measure $ \int_G {\rm d}\mu_G$ and corresponding plethystic exponential defined in \cref{def:PExp}, the
Hilbert series is computed by the following formula
\begin{equation}
g = \int_G {\rm d}\mu_G ~ {\rm PE} \left[ \chi^G_R (z_a) \sum_i^{N_f} t_i \right] ~,
\end{equation}
where $\chi_R^G(t_i, z_a)$ is the character for representation $R$ of group $G$ and it is expanded into monomials of complex variables
$z_i$ and the Haar measure is given by
\begin{equation}
\int_G {\rm d}\mu_G = \frac{1}{|W|}\prod_{i=1}^{{\rm rank}(G)} \int_{|z_i| = 1} \frac{dz_i}{2\pi i z_i} \prod_{\alpha \in \Delta}(1 - \prod_i^{{\rm rank}(G)} z_i^{\alpha_i}) ~,
\end{equation}
where $W$ is the Weyl group and $\alpha$ is a root, or the weight of the adjoint
representation such that $\alpha_i$ is the  $i$-th entry of the weight vector in the Dynkin basis.
\end{theorem}
The remainder of this paper will be to evaluate this integral explicitly, first for some warm-up cases, and ultimately for the MSSM itself.

\section{Warm-up Examples}\setall
\label{sec:example}
Our goal is to apply the technology introduced in \S\ref{sec:PP} to the MSSM, with gauge group $\cG = SU(3) \times SU(2) \times U(1)$.
Before doing so, let us warm up with two illustrative examples: (1) the SQCD sector and (2) a single Abelian gauge theory.
This will give us a more concrete understanding of all the previous definitions from physical side.

\subsection{sQCD}
Let us look at the example of SQCD with $N_c$ colours and $N_f$ flavours, but without superpotential \cite{Gray:2008yu}.
Here, the GIOs are symmetric combinations of quarks and anti-quarks, transforming in the bifundamental $[1,0,\dots,0;0,\dots,0,1]$
of $SU(N_f)_L \times SU(N_c)$ and the bifundamental $[1,0,\dots,0;0,\dots,0,1]$ of $SU(N_c) \times SU(N_f)_R$ respectively. 
This is the quark sector of the calculation which we are about to perform.

In the above, we have used the standard Young diagram for irreducible representation of $SU(n)$. 
Let $\lambda_i$ be the length of the $i$-th row
($1 \leq i \leq n-1$) and let $a_i = \lambda_i - \lambda_{i-1}$ be the differences of lengths of rows. In such notation, we have a
representation written as $[a_1,a_2,\dots,a_{n-1}]$, of length $n-1$. 
For example, $[1,0,\dots,0]$ represents the fundamental representation,
$[0,\dots,0,1]$ represents the anti-fundamental representation, and $[1,0,\dots,0,1]$ (the second 1 is at the $(n-1)$-th position
represents the adjoint representation. For a product gauge group $SU(n) \times SU(n)$, we use notation $[\dots;\dots]$ where
the $(n-1)$-tuple to the left of the semicolon is the representation for the left $SU(n)$, and vice versa on the right.
Finally, let us denote the character for the (anti) fundamental representation of $SU(N)$ as $\chi_{[0,\dots,1]}^{SU(N)}$ and 
$\chi_{[1,0,\dots,0]}^{SU(N)}$ respectively.

To use the Weyl-Molien formula, we need to introduce weights for elements in the Cartan
subgroup for different groups. 
We use $z_a$, $a =1,\dots, N_c-1$ for colour weights and $t_i,\tilde{t}_i$, $i=1,\dots,N_f$ for flavour weights. 
Therefore, the character for a quark becomes $\chi_{[1,0,\dots,0;0,\dots,0,1]}^{SU(N_f)_L \times SU(N_c)}(t_i,z_a)$
and that for an anti-quark is $\chi_{[1,0,\dots,0;0,\dots,0,1]}^{SU(N_c) \times SU(N_f)_R}(\tilde{t}_i,z_a)$.
We further introduce two more variables for counting number of quarks and anti-quarks $t$ and $\tilde{t}$ respectively. 
The plethystic exponential from \eqref{def:PE} is precisely the object which constructs symmetric products of quarks and anti-quarks:
\begin{align}
\nonumber
&{\rm PE}\left[
t \chi_{[1,0,\dots,0;0,\dots,0,1]}^{SU(N_f)_L \times SU(N_c)}(t_i,z_a) +
\tilde{t} \chi_{[1,0,\dots,0;0,\dots,0,1]}^{SU(N_c) \times SU(N_f)_R}(\tilde{t}_i,z_a)
\right]\\
&\equiv \text{exp} \left[ 
\sum_{k=0}^\infty \frac{1}{k}(t^k \chi_{[1,0,\dots,0;0,\dots,0,1]}^{SU(N_f)_L \times SU(N_c)}(t_i^k,z_a^k) + 
\sum_{k=0}^\infty \frac{1}{k}(\tilde{t}^k \chi_{[1,0,\dots,0;0,\dots,0,1]}^{SU(N_c) \times SU(N_f)_R}(\tilde{t}_i^k,z_a^k)\right] \ .
\end{align}
Expanding the character more explicitly as
\begin{equation}
t \chi_{[1,0,\dots,0;0,\dots,0,1]}^{SU(N_f)_L \times SU(N_c)}(t_i,z_a) = \chi_{[0,\dots,0,1]}^{SU(N_c)}(z_l) \sum_{i=1}^{N_f} t_i \quad,
\end{equation}
gives us
\begin{align}
\nonumber
&\text{PE} \left[
\chi_{[1,0,\dots,0]}^{SU(N_c)}(z_l) \sum_{i=1}^{N_f} \tilde{t}_i + \chi_{[0,\dots,0,1]}^{SU(N_c)}(z_l) \sum_{j=1}^{N_f} t_j
\right]\\
& = \text{exp} \left[
\sum_{k=0}^{\infty} \left( \frac{\chi_{[1,0,\dots,0]}^{SU(N_c)}(z_l^k) \sum_{i=1}^{N_f} \tilde{t}_i^k + \chi_{[0,\dots,0,1]}^{SU(N_c)}(z_l^k) \sum_{j=1}^{N_f} \tilde{t}_j^k}{k}
\right)
\right]\quad.
\end{align}
Here we associated dummy variables $t$ and $\tilde{t}$ to stand for quarks and anti-quarks counting the global $U(1)$ charges in
the maximal torus of the global symmetry. Therefore, we should restrict the values of $t_i$ to be $|t_i| < 1$ for all $i$.

As described in \S\ref{s:MW}, we want gauge invariant quantities, therefore, it is important that we project these representations onto trivial
subrepresentations that are made up by quantities \ep{invariant} under the action of gauge group. 
The Molien-Weyl integral from \eqref{def:molien} thus gives the requisite Hilbert series (generating function) for $(N_f,N_c)$ as
\begin{equation}
g^{(N_f,N_c)} = \int_{SU(N_c)} \textrm{d}\mu_{SU(N_c)} \textrm{PE}\left[ \chi_{[1,0,\dots,0]}^{SU(N_c)}(z_l) \sum_{i=1}^{N_f} 
\tilde{t}_i + \chi_{[0,\dots,0,1]}^{SU(N_c)}(z_l) \sum_{j=1}^{N_f} t_j \right] \quad .
\end{equation}
The Haar measure $\textrm{d}\mu_{SU(N_c)}$ can be explicitly written using Weyl's integration formula as (see, \ep{e.g.,} Sec. 26.2 of \cite{Fulton:1991rt})
\begin{equation}
\int_{SU(N_c)} \textrm{d}\mu_{SU(N_c)} = \frac{1}{(2 \pi i)^{N_c -1} N_c!} \oint_{|z_l|=1} \prod_{l=1}^{N_c -1}\frac{\textrm{d}z_l}{z_l}\Delta(\phi)\Delta(\phi^{-1}) \quad,
\end{equation}
where $\phi_a(z_1,\dots,z_{N_c-1})_{a=1}^{N_c}$ are the coordinates on the maximal torus of $SU(N_c)$ with $\prod_{a=1}^{N_c}\phi_a =1$, and 
$\Delta(\phi) = \prod\limits_{1\leq a\leq b \leq N_c}(\phi_a - \phi_b)$ is the Vandermonde determinant.

Finally, let us construct the characters in the plethystic exponential. 
First we take the weights of the fundamental representation of $SU(N_c)$ to be
\begin{equation}
L_1 = (1,0,\dots,0), \quad L_k(0,0,\dots,-1,1,\dots,0), \quad L_{N_c} = (0,\dots,-1) \quad,
\end{equation}
where all $L$'s are $(N_c-1)$-tuples, and $L_k$ ($2 \leq k \leq N_c-1$) has $-1$ in the $(k-1)$-th  position and 1 in the $k$-th position. 
With this particular choice of weights, the coordinates on the maximal torus of $SU(N_c)$ are given by
\begin{equation}
\phi_1 = z_1, \quad \phi_k = z_{k-1}^{-1}z_k, \quad \phi_{N_c} = z_{N_c-1}^{-1} \quad,
\end{equation}
with $2 \leq k \leq N_c-1$.
Hence, the characters of the fundamental and anti-fundamental representations are
\be
\begin{split}
\chi_{[1,0,\dots,0]}^{SU(N_c)} &= \sum_{a=1}^{N_c}\phi_a = z_1 + \sum_{k=1}^{N_c-1}\frac{z_k}{z_{k-1}} + \frac{1}{z_{N_c-1}} \quad, \\
\chi_{[0,0,\dots,1]}^{SU(N_c)} &= \sum_{a=1}^{N_c}\phi_a^{-1} = \frac{1}{z_1} + \sum_{k=1}^{N_c-1}\frac{z_{k-1}}{z_k} + z_{N_c-1} \quad
\end{split}
\ee

Thus, we have that \cite{Gray:2008yu}:
\begin{theorem}
The final expression for the Hilbert series for SQCD is the ordinary integral
\begin{align}
g^{(N_f, N_c)}(t_i, \tilde{t}_i) 
&= \frac{1}{(2 \pi i)^{N_c-1} N_c!} \oint_{|z_l|=1}  \prod_{l=1}^{N_c-1} \frac{ d z_l}{z_l} \Delta( \phi )\Delta (\phi^{-1}) \times \label{gSQCD} \\
&
\nn \mathrm{PE}\: \left[
\left( z_1 + \sum_{k=2}^{N_c-1} \frac {z_k} {z_{k-1} } + \frac {1} {z_{N_c-1} } \right)
\sum_{i=1}^{N_f} t_i +
\left( \frac{1} {z_1}  + \sum_{k=2}^{N_c-1} \frac {z_{k-1} } {z_k} + z_{N_c-1} \right)
\sum_{j=1}^{N_f} \tilde{t}_j  \right] \ .
\end{align}
\end{theorem}
Note that this a {\it refined} Hilbert series in the $2N_f$ variables $t_i$ an $\tilde{t}_i$.

\subsection{An Abelian Gauge Theory}
We have reviewed in the previous subsection, a rather formal and general example to elucidate the contents of Molien-Weyl formula for SQCD with $SU(N)$ gauge group. 
However, the spirit of the integral can be captured by a simple example using $U(1)$ without loss of
generality \cite{Lehman:2015via}. 
First, consider a single complex scalar field charged under a $U(1)$ symmetry,
 \ep{i.e}., $\phi \rightarrow e^{i\theta} \phi$,
$\phi^* \rightarrow e^{-i\theta}$. Clearly, the gauge invariants are now $(\phi\phi^*)^n$ and there is only one such operator for each $n$.
We can then define a formal series as
\begin{equation}
H = \sum_{n=1}^\infty c_n(\phi \phi^*)^n \quad,
\end{equation}
where $c_n = 1$ counts the number of different invariants of a given $n$ (since there is clearly only one per degree), so when expanded, it is
\begin{equation}
H = 1+ (\phi \phi^*) + (\phi \phi^*)^2 + (\phi \phi^*)^3 + \cdots \ .
\end{equation}
If $(\phi, \phi^*)$ are formally treated as numbers less than one, it is simply a geometric series in variables $(\phi, \phi^*)$,
\begin{equation}\label{HU1}
H(\phi, \phi^*) = \frac{1}{1 - \phi \phi^*} \quad.
\end{equation}
Here, we obtain a refined Hilbert series in two variables; because the field itself is a complex scalar, we can identify the field with its own corresponding fugacity.

Introducing another variable $\theta$, the same variable that parametrises the $U(1)$, \eqref{HU1} can be re-written as the integral 
\begin{equation}
H = \frac{1}{2\pi}\int_0^{2\pi} \frac{\textrm{d}\theta}{(1 - \phi e^{i\theta})(1 - \phi^* e^{-i\theta})} \quad.
\end{equation}
This re-parametrisation can be seen by series expanding $(1 - \phi e^{i\theta})^{-1}(1 - \theta^* e^{-i\theta})^{-1}$, which is
$(1+\phi e^{i\theta} + (\phi e^{i\theta})^2 + \cdots )(1+\phi^* e^{-i\theta} + (\phi^* e^{-i\theta})^2 + \cdots )$. By multiplying out and collect
terms according to powers of $e^{i\theta}$, we see that the terms that are free of $e^{i\theta}$ are exactly the formal series we started with,
\ep{i.e.}, $1+ (\phi \phi^*) + (\phi \phi^*)^2 + (\phi \phi^*)^3 + \cdots$. The terms with any number of factors of $e^{i\theta}$ or $e^{-i\theta}$ vanish
upon the $\theta$ integration.

Making substitution $z = e^{i\theta}$, the d$\theta$ integral becomes a contour integral around $|z| = 1$.
\begin{equation}
H = \frac{1}{2\pi i } \oint_{|z|=1} \frac{\textrm{d}z}{z}\frac{1}{(1-\phi z)(1-\phi^* z^{-1})} \quad .
\end{equation}
We can further reframe the second part of the integrand
\be\label{eg:hs}
\begin{split}
\frac{1}{(1-\phi z)(1-\phi^* z^{-1})} 
&= \textrm{exp} \left[ -\textrm{log}(1-\phi z) - \textrm{log} (1- \phi^*z^{-1}) \right] \\
&= \textrm{exp} \left[ \sum_{n=1}^\infty \frac{(\phi z)^n}{n} + \sum_{n=1}^{\infty} \frac{(\phi^* z^{-1})^n}{n} \right] 
= \textrm{PE} \left[ \phi z + \phi^* z^{-1}  \right]
\quad.
\end{split}
\ee

To understand the previous lines, let us expand the LHS of Eq.\ref{eg:hs}, with $\phi$ and $\phi^*$ being small complex numbers. To cubic order 
in both fields we have
\be\label{eq:scalar_hs}
\begin{split}
\frac{1}{(1-\phi z)(1-\phi^* z^{-1})} &= 1+ \phi \phi^* + (\phi \phi^*)^2 + (\phi \phi^*)^3 + \cdots 
+ z (\phi + \phi (\phi \phi^*) + \phi (\phi \phi^*)^2 + \cdots )\\
&+ z^2 (\phi^2 + \phi^2(\phi \phi^*) + \cdots) + z^3\phi^3 + \phi^{*3} z^{-3} + z^{-2}(\phi^{*2}+\phi^{*2}(\phi \phi^*) + \cdots)\\
&+z^{1}(\phi^* + \phi^*(\phi \phi^*) + \phi^*(\phi \phi^*)^2 + \cdots) \quad.
\end{split}
\ee
From this expansion, the terms with no factors of $z$ are the ones invariant under $U(1)$, which are picked out by the
contour integral. \cref{eq:scalar_hs} shows that we can obtain the series of charge $+1$ by multiplying $z^{-2}$ 
and that of charge $-2$ by multiplying $z$. These results follow from the fact that the expansion has already generated all possible
 combinations of $\phi$ and $\phi^*$. In doing so, we implicitly used the reasoning behind \eqref{def:molien}. There are two underlying
 concepts running parallel. 
 
First, Eq.\ref{eg:hs} generates all possible arrangements of the scalar fields, as graded by charge. 
Indeed, we see the natural emergence of 
 the \ep{plethystic exponential}, as the generating function of all symmetric combination of its argument. 
 Second, the integration over $d\theta = \frac{dz}{i z}$ is the integration over the group manifold $U(1)$. 
 This makes sense as we want group invariant quantities, 
 so we have to ``average" over group elements. When integrated over $d\theta$, any terms with non-trivial powers of $z = e^{i\theta}$ become
 integrals $d\theta e^{i n\theta}$ for some integer $n$. This is identically 0 since integral is from 0 to $2\pi$. Hence, terms with no powers of $z$
 remain and are $U(1)$ invariant.

\section{Molien-Weyl Integral for the MSSM}\label{sec:setup}\setall
In this section, we set up the scene for performing the Molien-Weyl integral to obtain the Hilbert series of MSSM. 
Then, we will use the result to interpret geometrical properties for the VMS for the MSSM.
We emphasize that the analysis will be done with the superpotential $W=0$.

The group $\cG$ under consideration for MSSM is, of course, the product gauge group
$SU(3) \times SU(2)_L \times U(1)_Y$. 
The corresponding character for product group is then also a product
for individual factor group, following from the very definition of a group character.
Indeed, for a given group $G$, we can associate any representation $R$ with a character $\chi_R : G \rightarrow \mathbb{C}$, where
the map is defined to be the trace of any group element $g$ in representation $R$. Under this definition, the
character for direct sum and products for representation is given by $\chi_{R_i \oplus R_j} = \chi_{R_i} + \chi_{R_j}$
and $\chi_{R_i \otimes R_j} = \chi_{R_i} \chi_{R_j}$.
Thus equipped, we simply need to input the particle contents with corresponding representation
for the product gauge group of MSSM, along with appropriate Haar measure for each factor group, to construct the
integral in Theorem \ref{MW}.

The particle content for MSSM are well known and recapitulated in Table \ref{tab:MSSM-Contents}.
\begin{table}[htb!]
\centering
\begin{tabular}{|c|c|c|c|}
\hline
Field & Multiplicity & Representation & SM Particle\\
\hline
$Q$ & 3 & $(\bf{3}, \bf{2})_{\frac{1}{6}}$ & Left-handed quark doublet\\
$U^C$ & 3 & $(\bar{\bf{3}}, \bf{1})_{-\frac{2}{3}}$ & Right-handed up-type anti-quark\\
$D^C$ & 3 & $(\bar{\bf{3}}, \bf{1})_{\frac{1}{3}}$ & Right-handed down-type anti-quark\\
$L$ & 3 & $(\bf{1}, \bf{2})_{-\frac{1}{2}}$ & Left-handed lepton doublet\\
$E^C$ & 3 & $(\bf{1}, \bf{1})_{1}$ & Right-handed anti-lepton doublet\\
$H_u$ & 1 & $(\bf{1}, \bf{2})_{\frac{1}{2}}$ & Higgs\\
$H_d$ & 1 & $(\bf{1}, \bf{2})_{-\frac{1}{2}}$ & Higgs\\
\hline
\end{tabular}
\caption{{\sf Minimal Supersymmetric Standard Model particle contents are given in the table, where
the Representation column entries give information how each particle transform under the product
gauge group. For example, the first row means quark $Q$ transforms in fundamental representation $\bf{3}$
of $SU(3)$, $\bf{2}$ of $SU(2)$ and has charge $\frac{1}{6}$ under $U(1)$. In addition, $\bar{\bf{3}}$ means
anti-fundamental representation.}}\label{tab:MSSM-Contents}
\end{table}
The characters for each factor group are taken from \cite{Hanany:2008sb} and the relevant ones are presented in Table \ref{tab:MSSM-chi}.
\begin{table}[htb!]
\centering
\begin{tabular}{|c|c|}
\hline
$SU(3)$ fundamental & $\chi_{\bf{3}}^{SU(3)}(z_a) = z_1 + \frac{z_2}{z_1} + \frac{1}{z_2}$\\
$SU(3)$ anti-fundamental & $\chi_{\bar{\bf{3}}}^{SU(3)} (z_a) = \frac{1}{z_1} + \frac{z_1}{z_2} +z_2$\\
$SU(2)$ (anti-) fundamental & $\chi_{\bar{\bf{2}}}^{SU(2)}(y) = \chi_{\bf{2}}^{SU(2)}(y) = y + \frac{1}{y}$\\
$U(1)$ & $\chi(x)_{Q}^{U(1)} (x) = x^Q$\\
\hline
\end{tabular}
\caption{{\sf The characters used for constructing Plethystic exponential of MSSM.}}\label{tab:MSSM-chi}
\end{table}
Finally, the Haar measures for each group \cite{Hanany:2008sb} are presented in Table \ref{tab:MSSM-Haar}.
\begin{table}[htb!]
\centering
\begin{tabular}{|c|c|}
\hline
Group & Haar Measure\\
\hline
$SU(3)$ & $\int_{SU(3)} {\rm d} \mu_{SU(3)} = \frac{1}{(2 \pi i)^2} \oint_{|z_1|=1} \frac{{\rm d}z_1}{z_1}  \oint_{|z_2|=1} \frac{{\rm d}z_2}{z_2} (1-z_1z_2)(1-\frac{z_1^2}{z_2})(1-\frac{z_2^2}{z_1})$\\
$SU(2)$ & $\int_{SU(2)} {\rm d} \mu_{SU(2)} = \frac{1}{2 \pi i} \oint_{|y|=1} \frac{{\rm d}y}{y} (1- y^2)$\\
$U(1)$ &  $\int_{U(1)} {\rm d} \mu_{U(1)} = \frac{1}{2 \pi i} \oint_{|x|=1} \frac{{\rm d}x}{x}$\\
\hline
\end{tabular}
\caption{{\sf The Haar measure of MSSM gauge groups. Note here the Haar measures are different from
those of \cite{Gray:2008yu,Hanany:2008kn}. The ones presented here only use positive roots and do not need
Weyl group renormalisation.}}\label{tab:MSSM-Haar}
\end{table}

With the above data, we now proceed to explicitly construct the Molien-Weyl integral in its Plethystic  form, which we will call PI, 
with full MSSM contents. 
This simply involves putting each chiral field into its correct representation and input its quantum number for associated character. 
To do this, we first tabulate the characters of the particle content within MSSM in \cref{tab:MSSM-characters}.
\begin{table}[h]
	\centering
	$\begin{array}{c|c|c}
	\hline
		\textrm{Particle} &\textrm{Label} & \textrm{Character under } SU(3) \times SU(2)_L \times U(1)_Y\\ \hline
		 \textrm{Left-handed quarks } &Q_i& x^{\frac{1}{6}} \left(y+\frac{1}{y}\right) \left(z_1+\frac{1}{z_2}+\frac{z_2}{z_1}\right) \\ \hline
		\textrm{Right-handed up-type anti-quarks} &u_i& x^{-2/3}\left(\frac{z_1}{z_2}+z_2+\frac{1}{z_1}\right) \\ \hline
		\textrm{Right-handed down-type anti-quark} &d_i & x^{1/3} \left(\frac{z_1}{z_2}+z_2+\frac{1}{z_1}\right) \\ \hline
		\textrm{Left-handed lepton}&L_i & x^{-1/2}\left(y+\frac{1}{y}\right)\\ \hline
		\textrm{Right-handed anti-lepton} &e_i & x \\ \hline
		\textrm{Up-type Higgs} & H_u & x^{1/2} \left(y+\frac{1}{y}\right)\\ \hline
		\textrm{Down-type Higgs} & H_d & x^{-1/2} \left(y+\frac{1}{y}\right) \\ \hline		
	\end{array}$
	\caption{{\sf The characters of the MSSM particles under the Standard Model gauge group $SU(3) \times SU(2)_L \times U(1)_Y$.}}
	\label{tab:MSSM-characters}
\end{table}
We can then use the formula of Plethystic exponential in \cref{def:PE}
to obtain the integrand. For example, the exponent of the Plethystic exponential for left-handed quarks is
\begin{equation}
\begin{split}
{\rm log}\,
\big[
&y^9 z_1^6 z_2^6
\big\{\left(y-Q_1 \sqrt[6]{x} z_1\right) \left(1-Q_1 \sqrt[6]{x} y z_1\right) \left(y-Q_2 \sqrt[6]{x} z_1\right) \left(1-Q_2 \sqrt[6]{x} y z_1\right) \left(y-Q_3 \sqrt[6]{x} z_1\right)\\
& \left(1-Q_3 \sqrt[6]{x} y z_1\right) \left(z_2-Q_1 \sqrt[6]{x} y\right) \left(z_2-Q_2 \sqrt[6]{x} y\right) \left(z_2-Q_3 \sqrt[6]{x} y\right) \left(y z_2-Q_1 \sqrt[6]{x}\right)\\
&\left(y z_2-Q_2 \sqrt[6]{x}\right) \left(y z_2-Q_3 \sqrt[6]{x}\right) \left(y z_1-Q_1 \sqrt[6]{x} z_2\right) \left(z_1-Q_1 \sqrt[6]{x} y z_2\right) \left(y z_1-Q_2 \sqrt[6]{x} z_2\right)\\
&\left(z_1-Q_2 \sqrt[6]{x} y z_2\right) \left(y z_1-Q_3 \sqrt[6]{x} z_2\right) \left(z_1-Q_3 \sqrt[6]{x} y z_2\right)
\big\}^{-1}
\big]
\end{split}~,
\end{equation}
which upon taking exponential gives us the argument inside the logarithm. The Plethystic exponentials for the rest of the particle content can be obtained in the similar fashion. Thus we have the following
proposition.
\begin{small}
\begin{proposition}\label{prop:PI}
The Hilbert series for the VMS of the MSSM (with zero superpotential) is given by
\begin{equation}\label{eq:PI}
{\rm PI} = \frac{1}{(2 \pi i)^4} \oint_{|x|=1} \oint_{|y|=1} \oint_{|z_1|=1} \oint_{|z_2|=1} 
	{\rm PE}(x,y,z_1,z_2,Q_i,L_i,u_i,d_i,e_i,H_u,H_d)~,
\end{equation}
where $i = 1,2,3$ and ${\rm PE} (x,y,z_1,z_2,Q_i,L_i,u_i,d_i,e_i,H_u,H_d)$ is given by
\begin{equation}\label{eq:PI}
\begin{split}
&{\rm PE} (x,y,z_1,z_2,Q_i,L_i,u_i,d_i,e_i,H_u,H_d) = \\
&- x^9 y^{13} \left(y^2-1\right) z_1^{10} \left(z_1^2-z_2\right) z_2^{10} \left(z_1 z_2-1\right) \left(z_1-z_2^2\right) \\
& 
\Big[
\left(y-H_u \sqrt{x}\right) \left(H_d y-\sqrt{x}\right) \left(\sqrt{x} y-H_d\right) \left(H_u \sqrt{x} y-1\right) \left(x e_1-1\right) \left(x e_2-1\right)\\
&\left(x e_3-1\right) \left(\sqrt{x} y-L_1\right) \left(y L_1-\sqrt{x}\right) \left(\sqrt{x} y-L_2\right) \left(y L_2-\sqrt{x}\right) \left(\sqrt{x} y-L_3\right) \left(y L_3-\sqrt{x}\right) \\
&\left(\sqrt[3]{x} d_1-z_1\right) \left(\sqrt[3]{x} d_2-z_1\right) \left(\sqrt[3]{x} d_3-z_1\right) \left(x^{2/3} z_1-u_1\right) \left(x^{2/3} z_1-u_2\right) \left(x^{2/3} z_1-u_3\right) \\
&\left(y-\sqrt[6]{x} Q_1 z_1\right) \left(\sqrt[6]{x} y Q_1 z_1-1\right) \left(y-\sqrt[6]{x} Q_2 z_1\right) \left(\sqrt[6]{x} y Q_2 z_1-1\right) \left(y-\sqrt[6]{x} Q_3 z_1\right) \\
&\left(\sqrt[6]{x} y Q_3 z_1-1\right) \left(\sqrt[6]{x} y Q_1-z_2\right) \left(\sqrt[6]{x} y Q_2-z_2\right) \left(\sqrt[6]{x} y Q_3-z_2\right) \left(\sqrt[3]{x} d_1 z_1-z_2\right) \\
&\left(\sqrt[3]{x} d_2 z_1-z_2\right) \left(\sqrt[3]{x} d_3 z_1-z_2\right) \left(x^{2/3} z_2-u_1 z_1\right) \left(x^{2/3} z_2-u_2 z_1\right) \left(x^{2/3} z_2-u_3 z_1\right) \\
&\left(y z_2-\sqrt[6]{x} Q_1\right) \left(y z_2-\sqrt[6]{x} Q_2\right) \left(y z_2-\sqrt[6]{x} Q_3\right) \left(\sqrt[3]{x} d_1 z_2-1\right) \left(\sqrt[3]{x} d_2 z_2-1\right)\\
& \left(\sqrt[3]{x} d_3 z_2-1\right) \left(y z_1-\sqrt[6]{x} Q_1 z_2\right) \left(\sqrt[6]{x} y Q_1 z_2-z_1\right) \left(y z_1-\sqrt[6]{x} Q_2 z_2\right) \left(\sqrt[6]{x} y Q_2 z_2-z_1\right) \\
&\left(y z_1-\sqrt[6]{x} Q_3 z_2\right) \left(\sqrt[6]{x} y Q_3 z_2-z_1\right) \left(x^{2/3}-u_1 z_2\right) \left(x^{2/3}-u_2 z_2\right) \left(x^{2/3}-u_3 z_2\right)
\Big]^{-1}
 ~,
\end{split}
\end{equation}
where the fugacities $t_i \in \{Q_i,L_i,u_i,d_i,e_i,H_u,H_d\}$ are taken to be $|t_i| < 1$ due to the fact that they count the $U(1)$ charges
inside the maximal torus of the global symmetries.
\end{proposition}
\end{small}
As can be seen, even though with the help of the Molien-Weyl formula, we have reduced the problem of computing the Hilbert series to an ordinary contour integral, the result is still a formidable integral, involving an integrand which is a rational function with 8 factors in the numerator and 49 factors in the denominator!
The remainder of this paper is concerned with simplifying this integral explicitly and obtaining geometrical information therefrom.
Moreover, we remark that there are fractional powers in the integration variables which might upset the reader: after all, the final answer is a Hilbert 
series, which must be a rational function. We will show in the ensuing section that all fractional powers actually cancel or disappear in the course of the integration, as is required.

\section{Obtaining the MSSM Hilbert Series}\setall
\label{sec:MSSM_HS}
In previous section, we constructed the contour integral to obtain Hilbert series of MSSM in \eqref{eq:PI}, which gives the generating function that counts
the gauge invariant operators. 
As we can see that there are 4 variables $(x,y,z_1,z_2)$ that need to be integrated over and the pole structure of the integrand is quite complicated. Therefore, it would be illuminating to record the detailed steps in performing the integration and we shall see that the subtleties in pole positions become important during the integration procedure. The full codes and results can be accessed form \cite{url}.

\subsection{Finding Poles and Residues}
The integration procedure can be carried out as follows.
The intermediate results are too complicated to be presented in the text, or even in an appendix, but is available at following \href{https://github.com/xiao-yan/MSSM_Plethystics}{link} for the repository.
\subsubsection{The $y$ Integral} 
We simply integrate over the variable $y$ over contour $|y| = 1$. 
More specifically, according to the Residue Theorem,
we calculate the residue for poles for $y$ inside the contour prescribed. The poles for $y$ are functions of other complex variables with variable
$x$ having fractional power due to $1/6$ $U(1)$ charge of left-handed quarks .
These fractional powers will become a main reason for the complexity of later parts of the integral. 
With the only requirement of fugacities with modulus small than 1, the positions of these poles are completely determined, \emph{i.e.,}
whether inside or outside the unit circle prescribed for variable $y$. 
In fact, there are only 14 such poles for $y$:
\begin{align}
\nn
\frac{H_d}{\sqrt{x}}\ , \quad \frac{L_1}{\sqrt{x}}\ , \quad \frac{L_2}{\sqrt{x}}\ , \quad \frac{L_3}{\sqrt{x}}\ , \quad H_u \sqrt{x}\ , \quad Q_1 \sqrt[6]{x} z_1\ , \quad Q_2 \sqrt[6]{x}z_1\\
Q_3\sqrt[6]{x} z_1\ , \quad \frac{Q_1 \sqrt[6]{x}}{z_2}\ , \quad \frac{Q_2\sqrt[6]{x}}{z_2}\ , \quad \frac{Q_3 \sqrt[6]{x}}{z_2}\ , \quad \frac{Q_1 \sqrt[6]{x} z_2}{z_2}\ , \quad \frac{Q_2 \sqrt[6]{x} z_2}{z_1}\ , \quad \frac{Q_3 \sqrt[6]{x} z_2}{z_1} \ .
\end{align}


For each of these 14 poles, the residue can be readily obtained.
Normally, we would sum these 14 separate residues, put them under the same denominator and cancel any common factors between the inal denominator and numerator. 
However, this direct approach is already beyond computer package such as {\sc Mathematica}. 
To get a taste of the complexities of the rational functions under discussion,  let us present 2 of the 14 residues, all of which are complicated rational functions of similar complexity (again, the reader is referred to the above url for the full expressions as well as the {\sc Mathematica} code).
The residue for pole at $y=H_d/\sqrt{x}$ is
\begin{equation}
\begin{split}
&-x^{11} H_d^{13} z_1^{10} \left(z_1^2-z_2\right) z_2^{10} \left(z_1 z_2-1\right) \left(z_1-z_2^2\right)
\Big[ \left(x e_1-1\right) \left(x e_2-1\right) \left(x e_3-1\right) \left(H_d-x H_u\right)\\
& \left(H_d H_u-1\right) \left(H_d-L_1\right) \left(H_d L_1-x\right) \left(H_d-L_2\right) \left(H_d L_2-x\right) \left(H_d-L_3\right) \left(H_d L_3-x\right) \left(\sqrt[3]{x} d_1-z_1\right)\\
& \left(\sqrt[3]{x} d_2-z_1\right) \left(\sqrt[3]{x} d_3-z_1\right) \left(x^{2/3} z_1-u_1\right) \left(x^{2/3} z_1-u_2\right) \left(x^{2/3} z_1-u_3\right) \left(H_d-x^{2/3} Q_1 z_1\right)\\
& \left(H_d Q_1 z_1-\sqrt[3]{x}\right) \left(H_d-x^{2/3} Q_2 z_1\right) \left(H_d Q_2 z_1-\sqrt[3]{x}\right) \left(H_d-x^{2/3} Q_3 z_1\right) \left(H_d Q_3 z_1-\sqrt[3]{x}\right)\\
& \left(\sqrt[3]{x} d_1 z_1-z_2\right) \left(\sqrt[3]{x} d_2 z_1-z_2\right) \left(\sqrt[3]{x} d_3 z_1-z_2\right) \left(H_d Q_1-\sqrt[3]{x} z_2\right) \left(H_d Q_2-\sqrt[3]{x} z_2\right)\\
& \left(H_d Q_3-\sqrt[3]{x} z_2\right) \left(x^{2/3} z_2-u_1 z_1\right) \left(x^{2/3} z_2-u_2 z_1\right) \left(x^{2/3} z_2-u_3 z_1\right) \left(\sqrt[3]{x} d_1 z_2-1\right)\\
& \left(\sqrt[3]{x} d_2 z_2-1\right) \left(\sqrt[3]{x} d_3 z_2-1\right) \left(H_d z_2-x^{2/3} Q_1\right) \left(H_d z_2-x^{2/3} Q_2\right) \left(H_d z_2-x^{2/3} Q_3\right)\\
& \left(H_d z_1-x^{2/3} Q_1 z_2\right) \left(H_d Q_1 z_2-\sqrt[3]{x} z_1\right) \left(H_d z_1-x^{2/3} Q_2 z_2\right) \left(H_d Q_2 z_2-\sqrt[3]{x} z_1\right)\\
& \left(H_d z_1-x^{2/3} Q_3 z_2\right) \left(H_d Q_3 z_2-\sqrt[3]{x} z_1\right) \left(x^{2/3}-u_1 z_2\right) \left(x^{2/3}-u_2 z_2\right) \left(x^{2/3}-u_3 z_2\right)
\Big]^{-1}~,
\end{split}
\end{equation}
and the residue for pole at $y=L_1/\sqrt{x}$ is 
\begin{equation}\label{eq:res_y}
\begin{split}
&x^{11} L_1^{13} z_1^{10} \left(z_1^2-z_2\right) z_2^{10} \left(z_1 z_2-1\right) \left(z_1-z_2^2\right)
\Big[\left(x e_1-1\right) \left(x e_2-1\right) \left(x e_3-1\right) \left(L_1-H_d\right)\\
& \left(L_1-x H_u\right) \left(H_d L_1-x\right) \left(H_u L_1-1\right) \left(L_1-L_2\right) \left(L_1 L_2-x\right) \left(L_1-L_3\right) \left(L_1 L_3-x\right) \left(\sqrt[3]{x} d_1-z_1\right)\\
& \left(\sqrt[3]{x} d_2-z_1\right) \left(\sqrt[3]{x} d_3-z_1\right) \left(x^{2/3} z_1-u_1\right) \left(x^{2/3} z_1-u_2\right) \left(x^{2/3} z_1-u_3\right) \left(L_1-x^{2/3} Q_1 z_1\right)\\
& \left(L_1 Q_1 z_1-\sqrt[3]{x}\right) \left(L_1-x^{2/3} Q_2 z_1\right) \left(L_1 Q_2 z_1-\sqrt[3]{x}\right) \left(L_1-x^{2/3} Q_3 z_1\right) \left(L_1 Q_3 z_1-\sqrt[3]{x}\right)\\
& \left(\sqrt[3]{x} d_1 z_1-z_2\right) \left(\sqrt[3]{x} d_2 z_1-z_2\right) \left(\sqrt[3]{x} d_3 z_1-z_2\right) \left(L_1 Q_1-\sqrt[3]{x} z_2\right) \left(L_1 Q_2-\sqrt[3]{x} z_2\right)\\
& \left(L_1 Q_3-\sqrt[3]{x} z_2\right) \left(x^{2/3} z_2-u_1 z_1\right) \left(x^{2/3} z_2-u_2 z_1\right) \left(x^{2/3} z_2-u_3 z_1\right) \left(\sqrt[3]{x} d_1 z_2-1\right)\\
& \left(\sqrt[3]{x} d_2 z_2-1\right) \left(\sqrt[3]{x} d_3 z_2-1\right) \left(L_1 z_2-x^{2/3} Q_1\right) \left(L_1 z_2-x^{2/3} Q_2\right) \left(L_1 z_2-x^{2/3} Q_3\right)\\
& \left(L_1 z_1-x^{2/3} Q_1 z_2\right) \left(L_1 Q_1 z_2-\sqrt[3]{x} z_1\right) \left(L_1 z_1-x^{2/3} Q_2 z_2\right) \left(L_1 Q_2 z_2-\sqrt[3]{x} z_1\right)\\
& \left(L_1 z_1-x^{2/3} Q_3 z_2\right) \left(L_1 Q_3 z_2-\sqrt[3]{x} z_1\right) \left(x^{2/3}-u_1 z_2\right) \left(x^{2/3}-u_2 z_2\right) \left(x^{2/3}-u_3 z_2\right)
\Big]^{-1}~.
\end{split}
\end{equation}
We can see from the above expressions that the denominators have over 40 terms and the current built-in functions from the likes {\sc Mathematica} have difficulties in finding the common denominator and summing over the numerators even for these 2 terms, let along summing over all 14.
The reason is that with 40 terms, when brought to the same denominator and expanded, we are confronted with $2^{40} \sim 10^{12}$ monomial terms; factoring a polynomial with this many terms is clearly hopeless.
It is remarkable that we could forge ahead and obtain a final answer, as we shall see.

\subsubsection{The $z_1$ Integral} 
To circumvent the issue of summing all 14 residues from the $y$ integral, we perform the contour integral separately for each of the 14 expressions.
Specifically, these 14 rational expressions give poles whose positions are not fixed by condition in \ref{eq:PI} with respect to the contour
$|z_1| = 1$, \emph{i.e.}, we do not know whether some on the poles are inside or outside the contour. 
This simply means the contour integral can not be performed with the data available to us. 
However, the plethystic integral for MSSM should lead to a definite result and this indeterminacy of pole positions
of the intermediate steps should be a result of redundancy of doing the 14 integrals separately and not combining them into one rational expression.
If we could combine to previous 14 expressions into a single rational function with common factors cancelled between top and bottom,
the terms that give indeterminate pole positions in the denominators should disappear.

Therefore, we can simply make a choice for the fugacities. The ultimate answer cannot depend on this choice by construction.
We take the following choice: 

\begin{equation}\label{eq:fuga-choice}
\begin{split}
L_1=\frac{49}{51},\quad L_2=\frac{23}{34}, \quad L_3=\frac{25}{34}, \quad Q_1=\frac{8}{51}, \quad Q_2=\frac{25}{102}, 
\quad Q_3=\frac{5}{102}\\ 
u_1=\frac{101}{102}, \quad u_2=\frac{91}{102}, \quad u_3=\frac{14}{17}, \quad d_1=\frac{27}{34}, 
\quad d_2=\frac{5}{34}, \quad d_3=\frac{1}{102}\\
e_1=\frac{2}{3}, \quad e_2=\frac{2}{51}, \quad e_3=\frac{11}{17}, \quad H_d=\frac{101}{102}, \quad H_u=\frac{4}{17}.
\end{split}
\end{equation}
At first sight, this choice of variable seems random and arbitrary. Indeed, this choice is made randomly by {\sc Mathematica} subject to the constraints
that all the fugacities have modulus smaller than 1. It is reasonable to argue that there are infinitely many such choices and one could possibly
obtain some other final results with other choice of fugacities. However, the final answer from this choice of fugacities shows that the denominator is in
Euler form of $\prod (1-t^n)^{a_n}$, the coefficients of numerators are all integer and the Taylor expansion of the Hilbert series that is a rational function, gives
positive integer coefficients. It is reassuring to see all the checks for a legitimate Hilbert series go through and it is also a satisfying check in the future works
for different choices of fugacities to arrive at the same final answer.

With this choice of variables at hand, we can fully determine whether a pole is inside or outside of the contour, thus we know whether the pole
should be included when residues are collected. For example, we have a pole for the first of the 14 expression as $z_1=\frac{L_2}{Q_1 x^{2/3}}$.
If we only use the condition from Eq. \ref{eq:PI}, we will not be able to decide this should be included in the residue or not. 
However, with the choice from
\ref{eq:fuga-choice}, it is clear this should be discarded since it is outside the contour of $|z_1| = 1$.

Using this choice of fugacities, we arrive at a total of 198 poles that are inside the contour for collecting residues.
We obtain the integral for each individual rational function using {\sc Mathematica} built-in function. 
After performing the 198 integrals, we clean up the results
to reduce the amount of work for later integrals. This is done by collecting the terms sharing the same denominator and combining them into a single term. After these procedures, we arrive at a total of 114 terms (again, available at the aforementioned URL) that need to be integrated separately.

\subsubsection{The $z_2$ Integral} 
Using the results from previous step, we proceed to perform these 114 integrals over variable $z_2$. Using the choice
of fugacities \eqref{eq:fuga-choice}, the number of poles found the be located inside the contour $|z_2| = 1$ is 1622. This amount of computation
requires under 1 hour to complete on a laptop/PC with 4 cores using {\sc Mathematica} built-in function. 
However, there are large amount of redundancies within these computation.
This can be seen that there are terms that simply cancel when we sum all the terms and the number of terms is
reduced to 838. In addition, we can use the same method in the previous paragraph to collect together
terms that share the same denominator. 
The number of terms is now reduced to 574 to enter our final integration over $x$, which is quite reasonable.

\subsubsection{The $x$ Integral} 
First, the number of poles that are inside contour $|x| = 1$ with fugacity choice \ref{eq:fuga-choice} is 3106.
We proceed normally with the integral as before. 
To clean up the redundancies within these results, we sum up all the terms so that some of the terms will just cancel as they are
simply negative of each other. 
In addition, terms sharing common denominator are collected.
Finally, we obtain a list of 1538 terms. One important aspect of these results comes from the fact that even we started with a plethystic
integral with fractional power $1/6$ in variable $x$, we still end up with all terms have integer exponents and coefficients.
However, the raw results of 1538 terms contain terms with fractional exponents in some variables.
Remarkably, {\it these fractional exponents combine into integer ones when summed up}. 
Of course, the final answer for the Hilbert series is a rational function and cannot contain any fractional powers.
These extraordinary cancellations give us confidence that we are indeed doing the right thing.

To get a flavour of these terms, we present two of these terms which combine to give integer exponents, viz.,
\begin{equation}\label{eq:ex-frac-1}
\begin{split}
&Q_3^{11} u_1^{13} u_2^7 u_3^6  \Big[ 
2 \left(Q_3-Q_1\right)\left(Q_3-Q_2\right) \left(u_1-u_2\right) \left(u_1-u_3\right){}^2 \left(u_2-u_3\right)
\left(d_1 \sqrt{u_1} \sqrt{u_2}-\sqrt{u_3}\right)\\
&\left(d_2 \sqrt{u_1} \sqrt{u_2}-\sqrt{u_3}\right) \left(d_3 \sqrt{u_1} \sqrt{u_2}-\sqrt{u_3}\right)
\left(d_1 \sqrt{u_1} \sqrt{u_3}-\sqrt{u_2}\right) \left(d_2 \sqrt{u_1} \sqrt{u_3}-\sqrt{u_2}\right)\\
&\left(d_3 \sqrt{u_1} \sqrt{u_3}-\sqrt{u_2}\right) \left(\sqrt{u_1}-d_1 \sqrt{u_2} \sqrt{u_3}\right)
\left(\sqrt{u_1}-d_2 \sqrt{u_2} \sqrt{u_3}\right) \left(\sqrt{u_1}-d_3 \sqrt{u_2} \sqrt{u_3}\right) \\
&\left(e_1 \sqrt{u_1} \sqrt{u_2} \sqrt{u_3}-1\right) \left(e_2 \sqrt{u_1} \sqrt{u_2} \sqrt{u_3}-1\right)
\left(e_3 \sqrt{u_1} \sqrt{u_2} \sqrt{u_3}-1\right) \left(Q_3 u_1-Q_1 u_2\right) \left(Q_3 u_1-Q_2 u_2\right)\\
&\left(Q_1 Q_3 \sqrt{u_1} \sqrt{u_2}-\sqrt{u_3}\right) \left(Q_2 Q_3 \sqrt{u_1} \sqrt{u_2}-\sqrt{u_3}\right)
\left(Q_3^2 \sqrt{u_1} \sqrt{u_2}-\sqrt{u_3}\right) \left(Q_1 Q_3 \sqrt{u_1} \sqrt{u_3}-\sqrt{u_2}\right)\\
&\left(Q_2 Q_3 \sqrt{u_1} \sqrt{u_3}-\sqrt{u_2}\right) \left(Q_3^2 \sqrt{u_1} \sqrt{u_3}-\sqrt{u_2}\right)
\left(Q_1 Q_3 u_1^{3/2}-\sqrt{u_2} \sqrt{u_3}\right) \left(Q_2 Q_3 u_1^{3/2}-\sqrt{u_2} \sqrt{u_3}\right) \\
&\left(Q_3 u_1-Q_1 u_3\right) \left(Q_3 u_1-Q_2 u_3\right) \left(Q_3 u_1-H_d\right) \left(H_d Q_3 \sqrt{u_1}-\sqrt{u_2} \sqrt{u_3}\right) \left(H_u \, Q_3 u_1-1\right)\\
&\left(Q_3 \sqrt{u_1}-H_u \sqrt{u_2} \sqrt{u_3}\right) \left(Q_3 u_1-L_1\right) \left(Q_3 u_1-L_2\right) 
\left(Q_3 u_1-L_3\right)\\
&\left(L_1 Q_3 \sqrt{u_1}-\sqrt{u_2} \sqrt{u_3}\right) \left(L_2 Q_3 \sqrt{u_1}-\sqrt{u_2} \sqrt{u_3}\right)
\left(L_3 Q_3 \sqrt{u_1}-\sqrt{u_2} \sqrt{u_3}\right) 
\Big]^{-1} \quad,
\end{split}
\end{equation}
and
\begin{equation}\label{eq:ex-frac-2}
\begin{split}
&Q_3^{11} u_1^{13} u_2^7 u_3^6 \\
& \Big[
2 \left(Q_3-Q_1\right) \left(Q_3-Q_2\right) \left(u_1-u_2\right) \left(u_1-u_3\right){}^2 \left(u_2-u_3\right) \left(d_1 \sqrt{u_1} \sqrt{u_2}+\sqrt{u_3}\right)\\
&\left(d_2 \sqrt{u_1} \sqrt{u_2}+\sqrt{u_3}\right) \left(d_3 \sqrt{u_1} \sqrt{u_2}+\sqrt{u_3}\right) \left(d_1 \sqrt{u_1} \sqrt{u_3}+\sqrt{u_2}\right) \left(d_2 \sqrt{u_1} \sqrt{u_3}+\sqrt{u_2}\right)\\
&\left(d_3 \sqrt{u_1} \sqrt{u_3}+\sqrt{u_2}\right) \left(d_1 \sqrt{u_2} \sqrt{u_3}+\sqrt{u_1}\right) \left(d_2 \sqrt{u_2} \sqrt{u_3}+\sqrt{u_1}\right) \left(d_3 \sqrt{u_2} \sqrt{u_3}+\sqrt{u_1}\right)\\
&\left(e_1 \sqrt{u_1} \sqrt{u_2} \sqrt{u_3}+1\right) \left(e_2 \sqrt{u_1} \sqrt{u_2} \sqrt{u_3}+1\right) \left(e_3 \sqrt{u_1} \sqrt{u_2} \sqrt{u_3}+1\right) \left(Q_3 u_1-Q_1 u_2\right)\\
&\left(Q_3 u_1-Q_2 u_2\right) \left(Q_1 Q_3 \sqrt{u_1} \sqrt{u_2}+\sqrt{u_3}\right) \left(Q_2 Q_3 \sqrt{u_1} \sqrt{u_2}+\sqrt{u_3}\right) \left(Q_3^2 \sqrt{u_1} \sqrt{u_2}+\sqrt{u_3}\right)\\
&\left(Q_1 Q_3 \sqrt{u_1} \sqrt{u_3}+\sqrt{u_2}\right) \left(Q_2 Q_3 \sqrt{u_1} \sqrt{u_3}+\sqrt{u_2}\right) \left(Q_3^2 \sqrt{u_1} \sqrt{u_3}+\sqrt{u_2}\right)\\
&\left(Q_1 Q_3 u_1^{3/2}+\sqrt{u_2} \sqrt{u_3}\right) \left(Q_2 Q_3 u_1^{3/2}+\sqrt{u_2} \sqrt{u_3}\right) \left(Q_3 u_1-Q_1 u_3\right) \left(Q_3 u_1-Q_2 u_3\right)\\
&\left(Q_3 u_1-H_d\right) \left(H_d Q_3 \sqrt{u_1}+\sqrt{u_2} \sqrt{u_3}\right) \left(H_u Q_3 u_1-1\right) \left(H_u \sqrt{u_2} \sqrt{u_3}+Q_3 \sqrt{u_1}\right)\\
&\left(Q_3 u_1-L_1\right) \left(Q_3 u_1-L_2\right) \left(Q_3 u_1-L_3\right) \left(L_1 Q_3 \sqrt{u_1}+\sqrt{u_2} \sqrt{u_3}\right) \left(L_2 Q_3 \sqrt{u_1}+\sqrt{u_2} \sqrt{u_3}\right) \\
& \left(L_3 Q_3 \sqrt{u_1}+\sqrt{u_2} \sqrt{u_3}\right) 
\Big]^{-1}
\quad .
\end{split}
\end{equation}
As one can see, both expressions are sprinkled with troubling terms involving $\sqrt{u_i}$ and $u_i^{3/2}$.
Summing the expression \eqref{eq:ex-frac-1} and \eqref{eq:ex-frac-2} gives the common
denominator to be
\begin{equation}\label{eq:ex-frac-denom}
\begin{split}
&
2 \left(Q_3-Q_1\right) \left(Q_3-Q_2\right) \left(u_1-u_2\right) \left(u_1-u_3\right){}^2 \left(u_2-u_3\right) \left(d_1^2 u_1 u_2-u_3\right) \left(d_2^2 u_1 u_2-u_3\right) \left(d_3^2 u_1 u_2-u_3\right)\\ &\left(d_1^2 u_1 u_3-u_2\right) \left(d_2^2 u_1 u_3-u_2\right) \left(d_3^2 u_1 u_3-u_2\right) \left(u_1-d_1^2 u_2 u_3\right) \left(u_1-d_2^2 u_2 u_3\right) \left(u_1-d_3^2 u_2 u_3\right)\\
&\left(e_1^2 u_1 u_2 u_3-1\right) \left(e_2^2 u_1 u_2 u_3-1\right) \left(e_3^2 u_1 u_2 u_3-1\right) \left(Q_3 u_1-Q_1 u_2\right) \left(Q_3 u_1-Q_2 u_2\right) \left(Q_1^2 Q_3^2 u_1 u_2-u_3\right)\\
&\left(Q_2^2 Q_3^2 u_1 u_2-u_3\right) \left(Q_3^4 u_1 u_2-u_3\right) \left(Q_3 u_1-Q_1 u_3\right) \left(Q_3 u_1-Q_2 u_3\right) \left(Q_1^2 Q_3^2 u_1 u_3-u_2\right)\\
&\left(Q_2^2 Q_3^2 u_1 u_3-u_2\right) \left(Q_3^4 u_1 u_3-u_2\right) \left(Q_1^2 Q_3^2 u_1^3-u_2 u_3\right) \left(Q_2^2 Q_3^2 u_1^3-u_2 u_3\right) \left(Q_3 u_1-H_d\right)\\
&\left(H_d^2 Q_3^2 u_1-u_2 u_3\right) \left(H_u Q_3 u_1-1\right) \left(Q_3^2 u_1-H_u^2 u_2 u_3\right) \left(Q_3 u_1-L_1\right) \left(Q_3 u_1-L_2\right)\\
&\left(Q_3 u_1-L_3\right) \left(L_1^2 Q_3^2 u_1-u_2 u_3\right) \left(L_2^2 Q_3^2 u_1-u_2 u_3\right) \left(L_3^2 Q_3^2 u_1-u_2 u_3\right) 
\quad ,
\end{split}
\end{equation}
and all fractional powers disappear!
It is indeed reassuring that of the 3106 terms, any term with a fractional power therein has exactly 1 partner which cancels it upon summation.
This is guaranteed by representation theory (characters) in the Molien-Weyl formula.
The numerator is expanded to check for integer coefficients and exponents (note that this expansion gives us over 4 millions terms). 
The integer criterion indeed checks out for this example and the first few terms are 
\begin{equation}
\begin{split}
&2 d_1 d_2 d_3 H_d L_1 L_2 L_3 Q_1^2 Q_2^2 Q_3^{22} u_2^{13} u_3^6 u_1^{23}+2 d_1 d_2 H_d L_1 L_2 L_3 Q_1^2 Q_2 Q_3^{21} u_2^{12} u_3^7 u_1^{22}\\
&+2 d_1 d_3 H_d L_1 L_2 L_3 Q_1^2 Q_2 Q_3^{21} u_2^{12} u_3^7 u_1^{22}+2 d_2 d_3 H_d L_1 L_2 L_3 Q_1^2 Q_2 Q_3^{21} u_2^{12} u_3^7 u_1^{22}\\
&+2 d_1 d_2 H_d L_1 L_2 L_3 Q_1^2 Q_2^2 Q_3^{20} u_2^{12} u_3^7 u_1^{22}+2 d_1 d_3 H_d L_1 L_2 L_3 Q_1^2 Q_2^2 Q_3^{20} u_2^{12} u_3^7 u_1^{22}\\
&+2 d_2 d_3 H_d L_1 L_2 L_3 Q_1^2 Q_2^2 Q_3^{20} u_2^{12} u_3^7 u_1^{22}+2 d_1 d_2 d_3 H_d L_1 L_2 L_3 Q_1 Q_2 Q_3^{20} u_2^{12} u_3^7 u_1^{22}\\
&+2 d_1 d_2 d_3 H_d L_1 L_2 L_3 Q_1 Q_2^2 Q_3^{19} u_2^{12} u_3^7 u_1^{22}+2 d_1 d_2 d_3 H_d L_1 L_2 L_3 Q_1^2 Q_2 Q_3^{19} u_2^{12} u_3^7 u_1^{22} + \dots
\end{split}
\end{equation}

\subsection{Unrefining the Hilbert Series}
After the previous section, we now have a list of 1538 rational expressions that should be combined into a single rational function, which is the Hilbert Series for MSSM. However, due to
the complexity of each rational expression, it is impossible to combine even two terms under a common denominator using common computer packages such as
\texttt{Mathematica} or \texttt{Mccaulay2}. To show the complexity of each term, we present an example below
\begin{equation}
\begin{split}
&H_d^{15} Q_1^{14} u_1^9
\Big[ \left(Q_1-Q_2\right){}^2 \left(Q_1-Q_3\right){}^2 \left(u_1-u_2\right){}^2 \left(u_1-u_3\right){}^2 (H_d H_u-1)\\
&\left(H_d-L_1\right) \left(H_d-L_2\right) \left(H_d-L_3\right) \left(H_d Q_1^3-1\right) \left(H_d Q_1^2 Q_2-1\right) \left(H_d Q_1^2 Q_3-1\right)\\
&\left(d_1 H_d Q_1-1\right){}^2 \left(d_2 H_d Q_1-1\right){}^2 \left(d_3 H_d Q_1-1\right){}^2 \left(H_d^2 Q_1 Q_2-u_1\right)\\
&\left(H_d^2 Q_1 Q_3-u_1\right) \left(H_d-Q_1 u_1\right) \left(H_d-Q_2 u_1\right){}^2 \left(H_d-Q_3 u_1\right){}^2\\
&\left(H_d^2 Q_1^2-u_2\right) \left(H_d^2 Q_1^2-u_3\right) \left(H_u Q_1 u_1-1\right) \left(L_1-Q_1 u_1\right) \left(L_2-Q_1 u_1\right)\\
&\left(L_3-Q_1 u_1\right) \left(H_d Q_1-d_1 u_1\right) \left(H_d Q_1-d_2 u_1\right) \left(H_d Q_1-d_3 u_1\right) \left(e_1 H_d Q_1 u_1-1\right)\\
&\left(e_2 H_d Q_1 u_1-1\right) \left(e_3 H_d Q_1 u_1-1\right)
\Big]^{-1} \\
&\\
&L_1^{15} Q_1^{14} u_1^9
\Big[ \left(L_1-L_2\right) \left(L_1-L_3\right) \left(Q_1-Q_2\right){}^2 \left(Q_1-Q_3\right){}^2 \left(u_1-u_2\right){}^2 \left(u_1-u_3\right){}^2\\
&\left(L_1-H_d\right) \left(H_u L_1-1\right) \left(L_1 Q_1^3-1\right) \left(L_1 Q_1^2 Q_2-1\right) \left(L_1 Q_1^2 Q_3-1\right) \left(d_1 L_1 Q_1-1\right){}^2\\
&\left(d_2 L_1 Q_1-1\right){}^2 \left(d_3 L_1 Q_1-1\right){}^2 \left(H_d-Q_1 u_1\right) \left(H_u Q_1 u_1-1\right) \left(L_1^2 Q_1 Q_2-u_1\right)\\
&\left(L_1^2 Q_1 Q_3-u_1\right) \left(L_1-Q_1 u_1\right) \left(L_2-Q_1 u_1\right) \left(L_3-Q_1 u_1\right) \left(L_1-Q_2 u_1\right){}^2 \left(L_1-Q_3 u_1\right){}^2\\
&\left(L_1^2 Q_1^2-u_2\right) \left(L_1^2 Q_1^2-u_3\right) \left(L_1 Q_1-d_1 u_1\right) \left(L_1 Q_1-d_2 u_1\right) \left(L_1 Q_1-d_3 u_1\right)\\
&\left(e_1 L_1 Q_1 u_1-1\right) \left(e_2 L_1 Q_1 u_1-1\right) \left(e_3 L_1 Q_1 u_1-1\right)
\Big]^{-1}~.
\end{split}
\end{equation}
As we can see, a typical rational expression has roughly 33 factors in the denominator, thus combining them implies finding Lowest Common Multiple
between each denominator with roughly 30 factors.
That is, we need to compute resultants between all pairs from 1538 multivariate polynomials each with about $2^{30} \simeq 10^{9}$ monomial terms,
rendering the process impractical on an average PC/Laptop. 
Nevertheless, each of the 1538 rational functions is, as seen from the above expression, is not too complicated.
Therefore, we have
\begin{proposition}\label{fullHS}
The multivariate, fully refined Hilbert series for the MSSM (without superpotential) is a sum of 1538 rational functions in 17 variables,
viz., $Q_i,L_i,u_i,d_i,e_i$ for $i=1,2,3$ as well as $H_u$ and $H_d$.
The full expressions are given in \href{https://github.com/xiao-yan/MSSM_Plethystics}{this} link.
\end{proposition}

It is difficult to extract geometrical information directly from this full Hilbert series.
Happily, we can ``unrefine'', i.e., force the Hilbert series to be uni-variate by setting all 18 variables to a single one, say $t$, but to different powers.
That is, the unrefinement is simply a substitution of variables by expression $t^\alpha$, where $t$ is also a fugacity and $\alpha$ is the weight for particular variable that is being substituted. 
This weight normally corresponds to some particular $U(1)$ charges and some particular choice should render the common denominator non-zero when unrefining (cf. \cite{Feng:2007ur}). 
This particular choice of weight we make is as follows:

\begin{table}[h]
\centering
$\begin{array}{|ccccccccccccccccc|}
\hline
 d_1 & d_2 & d_3 & e_1 & e_2 & e_3 & H_d & H_u & L_1 & L_2 & L_3 & Q_1 & Q_2 & Q_3 & u_1 & u_2 & u_3 \\
 \hline
 t^{512} & t^{256} & t^{128} & t^{512} & t^{512} & t^{512} & t^{512} & t^{512} & t^{256} & t^{128} & t^{64} & t^{32} & t^{16} & t^8 & t^4 & t^2 & t \\
 \hline
\end{array}$
\caption{{\sf The weights for unrefining the Hilbert Series of MSSM.}}
\label{tab:weights}
\end{table}

With this choice of weights, we are able to greatly simplify the expression in Proposition \ref{fullHS} to a single rational function with denominator being in Euler product form and numerator having integer coefficients and exponents, as well shall now see. The astute reader might argue that this choice of wrights seems arbitrary and we can as well make other choices with more straightforward physical implications. In fact, we have made weight choices that represent baryron numbers, lepton numbers etc and find that they all give 0 in the denominator, thus disqualifying themselves as legitimate weight choices. This particular choice is by far the most reasonable weights that do not give 0 in the denominator in the rational function of Hilbert series.

\subsection{Simplifying the Unrefined Hilbert Series}
After the unrefinement with weights in \cref{tab:weights}, the Hilbert series is simplified into a rational function, with a polynomial of degree 816,890 as the
numerator and a denominator with 994 factors of total degree 824,397. 
Note that the factors are already in Euler form and should correspond to GIOs which parametrize the VMS of the MSSM.
Even this rational function looks unmanageable at first sight, we can still extract useful
information out of it. Importantly, as a preliminary step, we need to perform a Taylor series in $t$ for the Hilbert series in order to know how many independent  generators there are in each degree. In terms of the supersymmetric gauge theory, this counts the number of independent 1/2-BPS single-trace operators at each $U(1)$ charge \cite{Benvenuti:2006qr,Feng:2007ur}. 
Doing so we obtain:
\begin{align}
\nn
H(t) &= 1+2 t^2+4 t^3+6 t^4+10 t^5+16 t^6+20 t^7+28 t^8+38 t^9+48 t^{10}+64 t^{11}\\
	&+84 t^{12}+104 t^{13}+134 t^{14}+168 t^{15}+202 t^{16}+250 t^{17}+304 t^{18}+360 t^{19}+436 t^{20}+\mathcal{O}\left(t^{21}\right) \;.
\end{align}
It is very assuring that all coefficients are non-negative, as it is a requirement in the series development of the Hilbert series 
(since it counts the number of independent monomials in the polynomial ring corresponding to the variety). 
This requires highly
non-trivial conspiracy between the numerator and denominator since each contains many terms with explicitly negative coefficients. 
Furthermore, the leading term is 1, as is also required. 
The coefficient list $\{ 1,0,2,4,6,10,16,20,28,38,48,64,84,104,134,168,202,250,304,360,436 \ldots \}$,
unfortunately does not resemble anything known in the literature.  It would be interesting indeed if this appeared in some combinatorial context.

It is important that we cover the details for simplifying the Hilbert series $H(t) = P(t)/Q(t)$ to a usable form. Now, since after adding up the various partial 
fractions from the Molien-Weyl integral, the numerator $P (t)$ is a polynomial of degree $816,890 = 2\cdot5\cdot81,689$ of no particularly apparent structure and 
the denominator $Q(t)$, one of degree $824,397 = 3\cdot7\cdot37\cdot1061$, it is already rather beyond conventional packages such as \texttt{Mathematica} 
to simplify in any way.
At least the denominator is already in Euler form \footnote{
As a technical aside, we have to ensure that there are no terms like $(1 + t^{w})$ in the product in the denominator since the Euler must have all terms strictly with the minus sign.
We can guarantee this by multiplying, each time a term such as $(1 + t^{w})$ appear, numerator and denominator by $(1 - t^{w})$
so as to contribute a legitimate $(1 - t^{2w})$ factor in the denominator.
This actually happens only thrice: $(1 + t^8)(1 + t^{16})(1 + t^{24})$ in our case.
}
consisting of 991 unique factors, ranging from 2 at degree 1 to 1 at degree 1664 in our weighting.
\footnote{
With this 991 we are in fact familiar.
There are 991 generators of gauge invariants to the MSSM \cite{Gray:2005sr,Gray:2006jb}
However, by more recent re-calculations, this number is slightly higher than the correct value \cite{He:2014oha}. 
But we shall see that after more simplification, the number of
factors in the denominator reduces to 445. Thus, at this stage, it seems to be a curious coincidence. 
}

We can present the full expression for the denominator in a compact form: by the array $w_1^{a_1}, w_2^{a_2}, \ldots$ we mean the polynomial $(1 - t^{w_1})^{a_1} \cdot (1 - t^{w_2})^{a_2} \ldots$.
In this notation, the denominator $Q(t)$ is given by
\begin{align*}\label{eq:euler-denom}
	&240 \ 282 \ 358 \ 376 \ 386 \ 392 \ 400 \ 408^2 \ 413 \ 417 \ 422 \ 423 \ 424 \ 427 \ 429 \ 432^2 \ 433 \ 434 \ 435 \ 437 \\
	&438 \ 441 \ 442 \ 443 \ 444 \ 445 \ 448 \ 461 \ 462 \ 463 \ 464^2 \ 465 \ 466 \ 469 \ 470 \ 471 \ 472^2 \ 473 \ 474 \ 475 \\
	&476^2 \ 478^2 \ 479^2 \ 480 \ 484 \ 485 \ 486 \ 488^2 \ 489 \ 490 \ 492 \ 494 \ 495^2 \ 496 \ 500 \ 503 \ 506^2 \ 514 \ 520 \ 528 \\
	&531^2 \ 533^3 \ 534^2 \ 537^2 \ 538^2 \ 539^2 \ 540^3 \ 541^3 \ 542^2 \ 545 \ 546^3 \ 547^3 \ 548^3 \ 549^2 \ 550^2 \ 552^3 \ 553^2 \ 554^3 \\ 
	&555^3 \ 556^3 \ 557^2 \ 558^3 \ 559 \ 560^2 \ 561^2 \ 562^3 \ 563^2 \ 564^3 \ 565^3 \ 566^3 \ 567 \ 568^3 \ 569 \ 570^2 \ 572^2 \ 573 \\
	&576 \ 578 \ 579^3 \ 580 \ 581^2 \ 582^2 \ 584^2 \ 585^3 \ 586^3 \ 587 \ 588^3 \ 590 \ 591 \ 592^3 \ 593^3 \ 594^3 \ 595 \ 596^3 \ 598 \ 599 \\
	&600^3 \ 601 \ 602^3 \ 604^3 \ 605 \ 606 \ 607 \ 608^2 \ 609^3 \ 610^3 \ 611 \ 612^3 \ 614 \ 615 \ 616^3 \ 620 \ 623 \ 624^2 \ 625 \ 628 \\
	&631 \ 632^2 \ 633 \ 637 \ 642 \ 643^3 \ 644 \ 645^2 \ 646^2 \ 648^2 \ 649^3 \ 650^3 \ 651 \ 654 \ 655 \ 656^3 \ 657^3 \ 658^3 \ 659 \ 662 \\
	&663 \ 664 \ 668 \ 669 \ 670 \ 671 \ 672^2 \ 673^3 \ 674^3 \ 675 \ 676^3 \ 678 \ 679 \ 684 \ 686 \ 687 \ 688 \ 696 \ 700 \ 702 \ 703 \\
	&704^3 \ 710 \ 718 \ 720 \ 721 \ 722^3 \ 723 \ 724^3 \ 726 \ 727 \ 728^3 \ 729 \ 730^3 \ 732^3 \ 736^2 \ 737 \ 738 \ 740 \ 744^3 \ 745 \\
	&746^3 \ 747 \ 748^3 \ 749 \ 750 \ 752^3 \ 753 \ 758 \ 760^3 \ 766 \ 767 \ 768 \ 769 \ 770 \ 771^3 \ 772 \ 773^3 \ 774^2 \ 776 \ 777^3 \\
	&778^3 \ 780^3 \ 781 \ 782 \ 784^3 \ 785^3 \ 786^3 \ 788^3 \ 790 \ 792^3 \ 794 \ 796 \ 797 \ 798 \ 799 \ 800^2 \ 801^3 \ 802^3 \ 803 \ 804^3 \\
	&805 \ 806 \ 807 \ 808^3 \ 810 \ 811 \ 812 \ 813 \ 814 \ 815 \ 816 \ 819 \ 821 \ 822 \ 824 \ 828 \ 830 \ 831 \ 832^3 \ 848 \ 856 \ 864 \\ 
	&872 \ 880 \ 888 \ 892 \ 894 \ 895 \ 896^3 \ 903 \ 912 \ 920 \ 934 \ 936 \ 959 \ 963 \ 965 \ 974 \ 977 \ 978^3 \ 980^3 \ 982 \ 984^3 \\
	&985 \ 986 \ 988 \ 991 \ 992 \ 994 \ 995 \ 996 \ 997 \ 999 \ 1000^3 \ 1001 \ 1002 \ 1004 \ 1007 \ 1008 \ 1011 \ 1012 \ 1013 \ 1014 \\
	&1015 \ 1016^3 \ 1020 \ 1022 \ 1023 \ 1024^2 \ 1026^2 \ 1027^3 \ 1028^2 \ 1029^3 \ 1030^3 \ 1031^3 \ 1032^2 \ 1033^3 \ 1034^3 \ 1035 \\
	&1036^3 \ 1037 \ 1038 \ 1039 \ 1040^3 \ 1041^3 \ 1042^3 \ 1043 \ 1044^4 \ 1045 \ 1046 \ 1047 \ 1048^3 \ 1050 \ 1051 \ 1052 \ 1054\\
	&1055 \ 1056^2 \ 1057^3 \ 1058^3 \ 1059 \ 1060^4 \ 1061 \ 1062 \ 1063 \ 1064^3 \ 1068 \ 1070 \ 1071 \ 1072^3 \ 1076 \ 1078 \ 1079 \\
	&1083 \ 1084 \ 1086 \ 1087 \ 1088^3 \ 1090 \ 1098 \ 1100 \ 1101 \ 1102 \ 1103 \ 1114 \ 1120 \ 1124 \ 1126 \ 1127 \ 1136 \ 1140 \\
	&1142 \ 1143 \ 1144 \ 1148 \ 1150 \ 1151 \ 1152^3 \ 1154 \ 1155 \ 1162 \ 1164 \ 1166 \ 1167 \ 1178 \ 1216 \ 1241 \ 1242 \ 1244 \\
	&1248 \ 1252 \ 1254 \ 1257 \ 1258 \ 1260 \ 1264 \ 1268 \ 1270 \ 1272 \ 1276 \ 1278 \ 1279 \ 1280^3 \ 1282 \ 1290 \ 1292 \ 1293 \\
	&1294 \ 1295 \ 1304^3 \ 1306 \ 1320^3 \ 1472 \ 1508 \ 1512^3 \ 1514 \ 1515 \ 1517 \ 1524 \ 1526 \ 1527 \ 1528^3 \ 1530 \ 1531 \\
	&1533 \ 1536 \ 1547 \ 1548 \ 1549 \ 1550 \ 1552^3 \ 1555 \ 1557 \ 1558 \ 1563 \ 1565 \ 1566 \ 1571 \ 1573 \ 1574 \ 1578 \ 1579 \\
	&1581 \ 1582 \ 1587 \ 1589 \ 1590 \ 1664
\end{align*}

We now need to extract as much of the list of factors in $Q(t)$ from the numerator $P(t)$.
First, we know this is going to be possible because we can check that $P(1)=0$ (even on Mathematica this is still doable as this is
simply the sum over all coefficients) so that it must divide $(1-t)$ at least once (and as we will see, many times).
To efficiently perform factorization, we will use a so-called {\bf extended synthetic division algorithm}\cite{Fan:2003} for mono-variate polynomials.
Luckily, there is an available Python/Sage implementation(c.f. Appendix \ref{append:extsyndiv} for a detailed discussion of this algorithm) of whose liberal use we will take advantage.

Our strategy is to first go over the 991 factors (with multiplicity this amounts to 1477 factors) of the form $1-t^a$ and try synthetic division into the numerator, this will cancel any such Euler factors therefrom.
Doing so (and even with Python, it still takes on the order of 3 days on a regular laptop due to the large degree of the dividend), we find that
the numerator now reduces to a polynomial $P_1(t)$, of degree  $816,890  - 259,498 = 557,392$, (significantly reduced from the 816,890 of $P(t)$), likewise, the denominator
reduces to $Q_1(t)$m with only 445 unique factors (and with multiplicity, 684 factors), in the shorthand notation for the Euler form, $Q_1(t)$ is 
\begin{equation}\label{eq:euler-denom}
{\tiny
\arraycolsep=0.8pt\def\arraystretch{1.1}
\begin{array}{cccccccccccccccccccc}
 240 & 282 & 358 & 376 & 386 & 392 & 400 & 408^2 & 413 & 417 & 422 & 423 & 424 & 427 & 429 & 432^2 & 433 & 434 & 435 & 437 \\
 438 & 441 & 442 & 443 & 444 & 445 & 448 & 461 & 462 & 463 & 464^2 & 465 & 466 & 469 & 470 & 471 & 472^2 & 473 & 474 & 475 \\
 476^2 & 478^2 & 479^2 & 480 & 484 & 485 & 486 & 488^2 & 489 & 490 & 492 & 494 & 495^2 & 496 & 500 & 503 & 506^2 & 514 & 520 & 528 \\
 531^2 & 533^3 & 534^2 & 537^2 & 538^2 & 539^2 & 540^3 & 541^3 & 542^2 & 545 & 546^3 & 547^3 & 548^3 & 549^2 & 550^2 & 552^3 & 553^2 & 554^3 & 555^3 & 556^3 \\
 557^2 & 558^3 & 559 & 560^2 & 561^2 & 562^3 & 563^2 & 564^3 & 565^3 & 566^3 & 567 & 568^3 & 569 & 570^2 & 572^2 & 573 & 576 & 578 & 579^3 & 580 \\
 581^2 & 582^2 & 584^2 & 585^3 & 586^3 & 587 & 588^3 & 590 & 591 & 592^3 & 593^3 & 594^3 & 595 & 596^3 & 598 & 599 & 600^3 & 601 & 602^3 & 604^3 \\
 605 & 606 & 607 & 608^2 & 609^3 & 610^3 & 611 & 612^3 & 614 & 615 & 616^3 & 620 & 623 & 624^2 & 625 & 628 & 631 & 632^2 & 633 & 637 \\
 642 & 643^3 & 644 & 645^2 & 646^2 & 648^2 & 649^3 & 650^3 & 651 & 654 & 655 & 656^3 & 657^3 & 658^3 & 659 & 662 & 663 & 664 & 668 & 669 \\
 670 & 671 & 672^2 & 673^3 & 674^3 & 675 & 676^3 & 678 & 679 & 684 & 686 & 687 & 688 & 696 & 700 & 702 & 703 & 704^3 & 710 & 718 \\
 720 & 721 & 722^3 & 723 & 724^3 & 726 & 727 & 728^3 & 729 & 730^3 & 732^3 & 736^2 & 737 & 738 & 740 & 744^3 & 745 & 746^3 & 747 & 748^3 \\
 749 & 750 & 752^3 & 753 & 758 & 760^3 & 766 & 767 & 768 & 769 & 770 & 771^3 & 772 & 773^3 & 774^2 & 776 & 777^3 & 778^3 & 780^3 & 781 \\
 782 & 784^3 & 785^3 & 786^3 & 788^3 & 790 & 792^3 & 794 & 796 & 797 & 798 & 799 & 800^2 & 801^3 & 802^3 & 803 & 804^3 & 805 & 806 & 807 \\
 808^3 & 810 & 811 & 812 & 813 & 814 & 815 & 816 & 819 & 821 & 822 & 824 & 828 & 830 & 831 & 832^3 & 848 & 856 & 864 & 872 \\
 880 & 888 & 892 & 894 & 895 & 896^3 & 903 & 912 & 920 & 934 & 936 & 959 & 963 & 965 & 974 & 977 & 978^3 & 980^3 & 982 & 984^3 \\
 985 & 986 & 988 & 991 & 992 & 994 & 995 & 996 & 997 & 999 & 1000^3 & 1001 & 1002 & 1004 & 1007 & 1008 & 1011 & 1012 & 1013 & 1014 \\
 1015 & 1016^3 & 1020 & 1022 & 1023 & 1024^2 & 1026^2 & 1027^3 & 1028^2 & 1029^3 & 1030^3 & 1031^3 & 1032^2 & 1033^3 & 1034^3 & 1035 & 1036^3 & 1037 & 1038 & 1039 \\
 1040^3 & 1041^3 & 1042^3 & 1043 & 1044^4 & 1045 & 1046 & 1047 & 1048^3 & 1050 & 1051 & 1052 & 1054 & 1055 & 1056^2 & 1057^3 & 1058^3 & 1059 & 1060^4 & 1061 \\
 1062 & 1063 & 1064^3 & 1068 & 1070 & 1071 & 1072^3 & 1076 & 1078 & 1079 & 1083 & 1084 & 1086 & 1087 & 1088^3 & 1090 & 1098 & 1100 & 1101 & 1102 \\
 1103 & 1114 & 1120 & 1124 & 1126 & 1127 & 1136 & 1140 & 1142 & 1143 & 1144 & 1148 & 1150 & 1151 & 1152^3 & 1154 & 1155 & 1162 & 1164 & 1166 \\
 1167 & 1178 & 1216 & 1241 & 1242 & 1244 & 1248 & 1252 & 1254 & 1257 & 1258 & 1260 & 1264 & 1268 & 1270 & 1272 & 1276 & 1278 & 1279 & 1280^3 \\
 1282 & 1290 & 1292 & 1293 & 1294 & 1295 & 1304^3 & 1306 & 1320^3 & 1472 & 1508 & 1512^3 & 1514 & 1515 & 1517 & 1524 & 1526 & 1527 & 1528^3 & 1530 \\
 1531 & 1533 & 1536 & 1547 & 1548 & 1549 & 1550 & 1552^3 & 1555 & 1557 & 1558 & 1563 & 1565 & 1566 & 1571 & 1573 & 1574 & 1578 & 1579 & 1581 \\
  1582 & 1587 & 1589 & 1590 & 1664 &  & & & & & &  & & &  & &  &  &  & 
\end{array}
}
\end{equation}

It is also worthwhile to show a few terms of the numerator after this simplification
\be
\begin{split}
&1+644 t+207692 t^2+44723872 t^3+7234295277 t^4+937600779818 t^5+101422033650142 t^6\\
&+9418304332530212 t^7+766466167260384451 t^8+55530492150394701928 t^9\\
&+3626455298524160306256 t^{10}+215630005089109137483320 t^{11} +11771050533831821348659441 t^{12}\\
&+ \cdots +102352219991766 t^{556741}+944789516589 t^{556742} +7278808875 t^{556743}+44930916 t^{556744}\\
&+208335 t^{556745}+645 t^{556746}+t^{556747} \;.
\end{split}
\ee
Using the simplified set of numerator and denominator, we should proceed to extract more information such as dimension and degree
of the variety. To do this, let us recall that there are two forms of Hilbert series in Eq. \ref{def:HS}. However, our HS has a different form in the
denominator due to our specific choice of weights. We can still obtain the dimension using the data at our disposal. The process is as follows
\begin{enumerate}
\item
Collect all the $(1-t)$ factors from the numerator, which in our case, the multiplicity is 644.
\item
Count the number of Euler factors of the form $(1-t^{m_i})$ with multiplicity, which is 684. The dimension is simply $688-644 = 40$.
\item
To get the degree, we put the HS into the form
\[\
{\rm HS} = \frac{(1-t)^{644}P_1(t)}{Q_1(t)} \;, \]\
where $Q(t)$ is in Euler form and $P_1(1) \neq 0$. Now the degree is $P_1(1)$ with factors of 2 being pulled out and ignored as they
come from our choice of weights. This is a rather large number about $-2.24\times10^{1633}$ and we present it in \cref{sec:intro}.
\end{enumerate}

Summarising what we have obtained so far, we start with the plethystic integral \cref{prop:PI} and 
perform the contour integral stepwise by finding residues of the integrand. Due to the complexities
of combining the intermediate results under the common denominator, we perform the ensuing integrations
separately as discussed after \cref{eq:res_y}. This brings extra redundancies during the integration
as it is not possible to determine whether some poles should be included in the integration since we
do not know if they are inside the contour or not. To eliminate these redundancies, we make some
choice of fugacities in \cref{eq:fuga-choice} and this choice is justified by the final result which
has Euler form in its denominator and is free from fractional powers and  coefficients. After obtaining
the final unrefined results, we are still presented with the problem of extracting useful geometric
information out of this complex expression. Unrefining the expression as suggested in \cref{tab:weights}
is thus necessary. Carrying out this procedure, we still have to use the extended synthetic division
algorithm in \cref{append:extsyndiv} to further simplify the rational function. Finally, the dimension
and degree are obtained by transforming the HS into the two forms of Hilbert series in \cref{thm:HS-dim-deg} and \cref{thm:HS-partial-deg}.
Finally, we obtain the dimension of the Hilbert series to be 40 and the degree as stated in 
\cref{append:detailres}.
With such results, we summarise them into the following theorem

\begin{theorem}
	The Hilbert series for MSSM with gauge group $SU(3) \times SU(2) \times U(1)$ and particle content
	in \cref{tab:MSSM-Contents} is a rational function with its numerator being a polynomial of degree
	557,392 and its denominator being in Euler form with structure shown in \cref{eq:euler-denom}. The
	detailed results for Hilbert series can be found in this \href{https://github.com/xiao-yan/MSSM_Plethystics}{link}. Particularly, the dimension
	obtained from the HS is 40 and the degree is shown in \cref{sec:intro}.
\end{theorem}

\section*{Acknowledgments}

YHH is indebted to the Science and Technology Facilities Council, UK, for grant ST/J00037X/1, the Chinese Ministry of Education,
for a Chang-Jiang Chair Professorship at NanKai University, and the city of Tian-Jin for a Qian-Ren Award. 
YX is grateful for the City Doctoral Scholarship for its generous support.

\appendix

\section{Illustrative Examples for the Plethystic Programme}

The first part of this appendix reviews the application of plethystics in converting between single- and multi-trace partition functions that count BPS operators.

\paragraph{Single-Trace at $N \rightarrow \infty$}
To familiarise ourselves with the definitions in \cref{def:PE}, we take $\mathbb{C}^3$ for illustration. 
This comes from the AdS/CFT correspondence where the CY threefold is simply $\mathbb{C}^3$ with associating Sasaki-Einstein manifold being $S^5$.
There is no baryonic charge since the third homology of $S^5$ is trivial and the isometry group is $SU(4)$ of rank 3, meaning CY manifold is toric and has 3
$U(1)$ charges.
So we can define 3 variables $t_1,t_2,t_3$ for measuring charges in their powers. The $\mathcal{N}=4$ $U(N)$
gauge theory in $\mathcal{N}=1$ language has three adjoint chiral multiplets $x,y$ and $z$. Since we want to count GIOs, we therefore
need to impose F-term relations from superpotential $W={\rm Tr}(x[y,z])$. By solving these constraints, we have the relation 
$[x,y] = [y,z] = [z,x] = 0$. Therefore, a generic single-trace GIO in the chiral ring will be of the form ${\rm Tr}(x^iy^jz^k)$. Then we can assign
$t_1$ to count the number of field $x$, $t_2$ the number of $y$ and $t_3$, the number of $z$. Putting together, we have the generating
function to be
\begin{equation}\label{eq:C3PE}
f(t_1,t_2,t_3;\mathbb{C}^3) = \sum_{i=0}^\infty \sum_{j=0}^\infty \sum_{k=0}^\infty t_1^i t_2^j t_3^k = \frac{1}{(1-t_1)(1-t_2)(1-t_3)} \quad.
\end{equation}
This result becomes exact when we take the limit $N \rightarrow \infty$.
In the next paragraph, we direct out attention to the
relation between single- and multi-trace generating
function via plethystic exponential. 
\paragraph{Multi-Trace at $N \rightarrow \infty$} For the case of a single D3-brane on $\C^3$, the adjoint fields
$x,y$ and $z$ are simple complex numbers and thus any product of these fields are multi-trace operators. Therefore, we only have
4 single trace operators: the identity, $x,y$ and $z$. So the generating function for single-trace becomes 
\[\ f_1(t_1,t_2,t_3) = 1 + t_1 + t_2 + t_3\]\ 
Now we look at the single-trace generating function for $N \rightarrow \infty$, which is \cref{eq:C3PE}. Each of such operators
is represented by a monomial $t_1^it_2^jt_3^k$, which can be interpreted as a multi-trace operator for just $N=1$ or one D3-brane.
Therefore, this means $g_1$, the generating function for multi-trace operators on a single D3-brane is the same as $f_\infty$, the
generating function for single-trace operators for infinite D3 branes: $g_1 = f_\infty$. Now we find the relation between
$f_1$ and $g_1$:
\begin{equation}\label{eq:PE-derive}
\begin{split}
g_1(t_1,t_2,t_3) & = \frac{1}{(1-t_1)(1-t_2)(1-t_3)} = {\rm exp}[-{\rm log}(1-t_1)-{\rm log}(1-t_2)-{\rm log}(1-t_3)]\\
&= {\rm exp}\left(\sum_{r=1}^\infty \frac{t_1^r+t_2^r+t_3^r}{r}\right) = {\rm exp}\left(\sum_{r=1}^\infty \frac{f_1(t_1^r,t_2^r,t_3^r)-1}{r}\right) \quad.
\end{split}
\end{equation}
We can see from the above that the function $g_1$ is the Plethystic Exponential of $f_1$ and this relation in fact generalise to any value
of $N$.

After this short review on plethystic exponential, we see that it is a combinatoric tool for generating
the Hilbert series or simply a generating function of all symmetric combination of its argument. It is interesting to see that the inverse of the exponential also contains certain geometric information as we shall shortly cover. The definition of plethystic logarithm is as follows:
\begin{equation}
f(t) = PE^{-1}(g(t)) = \sum_{k=1}^{\infty}\frac{\mu(k)}{k}\log(g(t^k))~,
\end{equation}
where $\mu(k)$ is the M\"obius function
\begin{equation}
\mu(k) = 
\begin{cases}
0 & \text{$k$ has one or more repeated prime factors~,}\\
1 & k=1~,\\
(-1)^n & \text{$k$ is a product of $n$ distinct primes~.}
\end{cases}
\end{equation}
To illustrate the reverse of the plethystic exponential, \ep{i.e} the \ep{plethystic logarithm}, we use two examples: (1) the simplest non-abelian subgroup of $SU(3)$ Valentiner group, $\Delta(3\cdot3^2)$ and (2) the simple abelian $\mathbb{Z}_3$,
to illustrate how we can obtain the information on the generators and syzygies thereof for $\mathbb{C}^3/\Delta(3\cdot3^2)$ and $\mathbb{C}^3/\mathbb{Z}^3$ as we as the determination of whether these two orbifolds are complete intersections. 
Consider the simplest non-Abelian discrete subgroup of $SU(3)$, \ep{i.e.}, the Valentiner group $\Delta(3\cdot3^2)$,
defined as
\setcounter{footnote}{0}
\be
\Delta(27) :=\braket{ \begin{pmatrix}\omega_3 & 0 & 0\\0 & 1 & 0 \\0 & 0& \omega_3^{-1} \end{pmatrix},
\begin{pmatrix}1 & 0 & 0\\0 & \omega_3 & 0 \\0 & 0& \omega_3^{-1} \end{pmatrix},
\begin{pmatrix}0 & 1 & 0\\0 & 0 & 1 \\1 & 0& 0 \end{pmatrix}} \quad.
\ee
The Molien series is given by
\[\
M(t;\Delta(27)) = \frac{-1 + t^3 - t^6}{(-1+t^3)^3} = 1 + 2t^3 + 4t^6 + 7 t^9 +11t^{12} + 16t^{15} + 22t^{18} + \cdots
\]\
First we need to construct its invariant generators and syzygies using a technique from Reynolds and Gr\"{o}bner basis.
Then we can check these results against those from plethystic logarithm.

The defining equations (syzygies), are constrained by the order of the group $\Delta(3\cdot3^2)$ \footnote{Specifically, this is a theorem due to N\"{o}ther: 
\ep{The polynomial ring of invariants is finitely generated and the degree of the generators is bounded by the order of the group $|G|$.}}
and we can construct this finite set of invariants. There is an averaging technique due to O. Reynolds (c.f. \cite{Yau:1993gq}. Given some polynomial
$F(x)$, one can construct the \ep{Reynolds operator}
\[\
R_G[F(x)] := \frac{1}{|g|}\sum_{g\in G} F(g \circ x) \quad.
\]\
Then the polynomial $R_G[F(x)]$ is invariant under $G$ by construction. Therefore, we go up to degree 27 to find the list of invariants for group $\Delta(3\cdot3^2)$.
More specifically, there are 174 invariants of degrees 0,3,6,...,24,27. With Gr\"{o}bner basis, we find that there are only 4 non-trivial generators for there
174 polynomials:
\[\
\{ m = 2xyz, \quad n = x^3 + y^3 + z^3, \quad p = x^6+ y^6+z^6, \quad q = x^3y^6 + x^6z^3 + y^3 z^6 \}.
\]\
We also find a single relation in $\mathbb{C}[m,n,p,q]$:
\be\label{eq:delta27}
8 m^6 + m^3(-48n^3 + 72np +72q) + 81((n^2-p)^3 -4n(n^2-p)q + 8q^2) = 0 \quad.
\ee
So we find that $\mathbb{C}^3/\Delta(3\cdot3^2)$ is a complete intersection given by a single hypersurface in $\mathbb{C}^4$.

Let us turn to plethystic logarithm, we find it for $\Delta(3\cdot3^2)$ to be
\[\
f_1 = PE^{-1} \left( \frac{-1+t^3-t^6}{(-1+t^3)^3} \right) = 2 t^3 + t^6 + t^9 - t^{18} \;.
\]\
We see that the RHS terminates and it can be interpreted as follows: there are 2 degree 3 invariants, 1 degree 6 and 1 degree 9 invariant, these 4 invariants
obey a single relation of total degree 18. Comparing this with \cref{eq:delta27}, we indeed see that this is the defining relation for $\mathbb{C}^3/\Delta(3\cdot3^2)$.
In fact, the finiteness of plethystic logarithm indicates that the underlying variety is a complete intersection, \ep{i.e.} the number of defining equation is equal
to the codimension of the variety in the embedding space. The story for non-complete intersection has more content to it. Now let us look at the abelian
orbifold $\mathbb{C}^3/\mathbb{Z}^3$, which is toric and also $dP_0$ as a cone over $\mathbb{P}^2$. For the group action $(x,y,z) \rightarrow \omega_3(x,y,z)$,
we can construct the Molien series to be
\be
f_\infty(t) = M(t,\mathbb{Z}_3) = \frac{1+7t^3+t^6}{(1-t^3)^3}\;,
\ee
where we can get the plethystic logarithm to be
\be
f_1(t) = PE^{-1} [f_\infty (t)] = 10 t^3 - 27 t^6 + 105 t^9 - 540 t^{12} + 3024 t ^{15} + \mathcal{O}(t^{18}) \;.
\ee
This agrees with known facts that the equation for this orbifold is 27 quadrics in $\mathbb{C}^{10}$, that is 10 degree three invariants satisfying 27 relations
of degree 6. However, these information are only included in the first two terms in the series and the rest of the terms are a reflection of the fact that we no
longer have a complete intersection. Therefore, the plethystic logarithm of the Hilbert series is no longer a polynomial and continues \ep{ad infinitum}.
In this final paragraph, let us explain why the plethystic logarithms for non-complete intersections are infinite. Firstly, the Poincar\'e series is always a rational
function when simplified and collected. Particularly, the denominator of the series is of the form of products of $(1-t^k)$ with possible repeats of $k$ while
the numerator being some complicated polynomial. We call this the \ep{Euler form}. When taking plethystic logarithm, we are essentially trying to solve the
following problem: find integers $b_n$ such that
\[\
f(t) = \frac{1}{\prod_{n=1}^\infty (1-t^n)^{b_n}}\;,
\]\
where $f(t)$ is a rational function in Euler form. Note that $PE^{-1}[f(t)] = \sum_{n=1}^\infty b_n t^n$ does not need to have all positive $b_n$. Since the
denominator is already in form of products of $(1-t^n)$, positive values of $n$ and $b_n$ can be read off immediately. The numerator in the rational function
gives the negative values of $b_n$ and contribute to the relations among invariants. For example, we can find $b_n$ for $\Delta(3\cdot3^2)$
\[\
\frac{1-t^3+t^6}{(1-t^3)^3} = \frac{(1-t^{18})}{(1-t^6)(1-t^9)(1-t^3)^2} = \frac{1}{\prod_{n=1}^\infty (1-t^n)^{b_n}}\;,
\]\
where we used the identity
\[\
\frac{(1-t^3)(1-t^{18})}{(1-t^6)(1-t^9)} = 1-t^3 + t^6\;.
\]\
Now we find the solution: the denominator contributes terms $2t^3, \;t^6$ and $t^9$ and the numerator contributes the terms $-t^{18}$. Thus,
$PE^{-1}[M(t)] = 2t^3 + t^6 + t^9 - t^{18}$. This means there are 2 degree 3, 1 degree 6 and 1 degree 9 invariants, obeying a single degree 18
relation. The crucial fact that the numerator can be factorized into Euler form dictates that the plethystic logarithm terminates in a series expansion.
Therefore, \ep{finding relation in this language corresponds to finding factorizations of the numerator into Euler form}. Take $\mathbb{C}^3/\mathbb{Z}_3$,
we have its Poincar\'e series as $(1-7t^3+t^6)/(1-t^3)^3$. No rational identity can put the numerator $1-7t^3+t^6$ into Euler form and the plethystic logarithm
does not terminate. If we convert the numerator into Euler form, we get
\be\label{eq:euler_exp}
1 +7t^3 +t^6 = \frac{(1-t^6)^{27}(1-t^{12})^{540}\cdots}{(1-t^3)^7(1-t^9)^{105}\cdots}\;,
\ee
where we have $10t^3$ from $(1-t^3)^{10}$ and $-27t^6$ from $(1-t^6)^{27}$. However, for higher degree invariants, 
\ep{i.e.}, 28 degree 6 and 55 degree 9 invariants, etc, we need  further expansion on both top and bottom
for \cref{eq:euler_exp}. Using computer package such as \texttt{Mccaulay2}, we can find 595 relations for 10 degree 3 and 28 degree 6 invariants:
55 of degree 6, 225 of degree 9 and 315 of degree 12. This thus reads
\[\
10 t ^3 + 28 t^6 -55 t^6 -225t^9 -315t^{12} = 10t^3 - 27t^6 - 225t^9 - 315t^{12}\;.
\]\
For higher degree invariants and relations, we can correct the coefficients for higher order terms such as $t^9$ and $t^{12}$.

\section{Extended Synthetic Division}\label{append:extsyndiv}
In this section, we review some basic materials of {\bf Extended Synthetic Division} with the \texttt{Python} implementation codes presented. Synthetic division
is a method of performing Euclidean division of polynomials with less calculation than regular polynomial long division. It is first developed for division
by monomial of the form $x-a$, but later generalized to division by any monomials and polynomials. The advantage of this method is that it allows one
to calculate division without writing out variables and it uses less calculations. Let us first look at a simple example:

\[\ \frac{x^3-12x^2-42}{x^2+x-3}\;. \]\
The steps are as follows
\begin{enumerate}
\item We negate them as before and write every coefficients but the first on to the left of the bar in an upward.
\item We copy the first coefficient and multiply the diagonal by the copied number and place them diagonally to the right from the copied entry.
\item We sum up the next column until we go past the entries at the top with the next diagonal multiplication.
\item We sum up the remaining column. Since there are two entries to the left of the bar, so the remainder is of degree 1. We then mark the separation with a vertical bar as
\[\ 1x \quad -13 \quad | \quad 16x \quad -81\quad, \]\
so we have the final quotient and remainder as
\[\
\frac{x^3-12x^2-42}{x^2+x-3} = x -13 + \frac{16x -81}{x^2+x-3} \quad.
\]\
\end{enumerate}
We present the division here for the conveinence of the reader
\begin{equation*}
{\tiny 
\centering
\begin{array}{ccccccc}
\begin{array}{cc|cccc}
& &1 & -12 & 0 & -42 \\ 
&3&   &    &    & \\
-1& & & & & \\ \cline{3-6}
\end{array} & \longrightarrow &
\begin{array}{cc|cccc}
& &1 & -12 & 0 & -42 \\ 
&3&   &    &   3 & \\
-1& & & -1& & \\ \cline{3-6}
&&1&&&
\end{array} \longrightarrow &
\begin{array}{cc|cccc}
& &1 & -12 & 0 & -42 \\ 
&3&   &    &   3 & -39\\
-1& & & -1& 13& \\ \cline{3-6}
&&1&-13&16&
\end{array} & \longrightarrow &\\
&&
\begin{array}{cc|cccc}
& &1 & -12 & 0 & -42 \\ 
&3&   &    &   3 & -39\\
-1& & & -1& 13& \\ \cline{3-6}
&&1&-13&16&-81
\end{array}&&&
\end{array}
}
\end{equation*}

More specifically, we present a \texttt{Python} implementation of the algorithm here
\begin{lstlisting}[language=python]
def extended_synthetic_division(dividend, divisor):

    out = list(dividend)
    normalizer = divisor[0]
    for i in xrange(len(dividend)-(len(divisor)-1)):
        out[i] /= normalizer 
        coef = out[i]
        if coef != 0:
            for j in xrange(1, len(divisor)): 
    separator = -(len(divisor)-1)
    return out[:separator], out[separator:]
\end{lstlisting}

\section{Hilbert Series, Gr\"obner Basis and Examples from Commutative Algebra}\label{append:com_alg}
In this appendix, we review some of the foundation of Hilbert series and see how it is constructed for specific counting purposes.
Firstly, we are most interested in {\bf polynomial ring} $K[x_1,\dots,x_n]$ consisting polynomials in variables $x_1,\dots,x_n$
with coefficients in the ring $K$. We typically take $K$ to be a field, such as real numbers $\mathbb{R}$. We also have monomials
in the form $x_1^{\alpha_1}\cdots x_n^{\alpha_n}$, whose linear combination gives a polynomials. So monomials serve as building blocks
for polynomials via addition. Since we are ultimately interested in counting things in polynomial ring using Hilbert Series, we would like to
simplify this counting to monomial level. Therefore, the notion of \emph{grading} is introduced for this counting purpose. On the physical
side story, the grading is usually from the charges of certain global symmetries. Let us look at
some natural choice for grading, the degree of a polynomial, defined as deg$(x_1^{\alpha_1}\cdots x_n^{\alpha_n}) = \alpha_1 + \cdots
+ \alpha_n.$ Adding up monomials of the same degree gives us a \ep{homogeneous} polynomial. Using this notation, we can decompose
a set of all homogeneous, degree $k$ polynomials $R_k$ into direct sum $R = \bigoplus_{k\in \mathbb{N}} R_k$. Mathematically, variables
$x_1,\dots,x_n$ are said to form a $\mathbb{N}$ graded algebra.

The dimension of $R_k$ is defined to be the number of independent degree $k$ monomials. A {\bf Hilbert Function} is
defined as $HF(R,k) = \text{dim}(R_k)$. The {\bf Hilbert series} is then naturally defined as 
\be
H(R,t) = \sum_k HF(R,k)t^k \quad.
\ee
For polynomial ring $R = K[x_1,...,x_n]$, to construct a degree $k$ monomial, we need to choose $k$ items from $n$ candidates, with
multiples being allowed, \ep{i.e.}
\be
HF(K[x_1,...,x_n], k) = \sum_{k_1+k_2+\cdots+k_n = k} \left(\begin{array}{c}n+k-1\\k\end{array}\right) \quad,
\ee
and the {\bf Hilbert series} is
\be\label{def:HS2}
H(K[x_1,...,x_n],k) = \sum_{k=0}^\infty \left( \begin{array}{c} n+k-1\\k \end{array}\right) t^k = \frac{1}{(1-t)^n}\quad.
\ee
The power of the denominator actually shows that there are $n$ degree 1 generators with no relations among them. Note that
this identity also gives the generating function of multiset coefficients.


Now let us discuss a bit more about ideal in a polynomial ring. First take $s$ polynomials from the ring, $f_1,f_2,...,f_s \in K[x_1,...,x_n]$.
Then the {\bf variety $V$} defined by $f_i$ are the points in $K^n$, which are zeroes of the $s$ polynomials. The ideal $\left<f_1,...,f_s\right>$
is then the set of polynomials that vanishes on $V$. With this definition in hand, we can proceed to define quotient variety now. 
Let $R = K[x_1,...,x_n]$ be a polynomial ring graded by degree and let $I = \left<f_1,...,f_s\right>$ be an ideal of $R$. Formally, the ideal
formed by polynomials $f_1,...f_s \in K[x_1,...,x_n]$ is the set of polynomials obtained by taking $f_i$ as basis vectors, with coefficients $h_i$
from $K[x_1,...x_n]$
\[\ \braket{f_1,...,f_s} = \left\{ \sum_{i=1}^s h_if_i : h_1,...,h_s \in K[x_1,...,x_n] \right\} \quad.\]\

We can
quotient the ring by the ideal,
\[\ M = R/I\quad. \]\
So by definition, $M$ is made of equivalence classes of polynomials. So in this sense,
the elements of ideal are zero polynomials and are removed by this quotient procedure, where the algebraic structure is preserve as
$M$ is also a ring. For homogenous ideal, the quotient ring derived from it is then defined to be a graded module since the grading is
understood to be the degree of polynomials. This means the quotient preserves the grading and $M$ is decomposed as a direct sum
$M = \oplus _k M_k$, where $M_k$ is the set of homogenous polynomials. So the Hilbert function for $M$ 
is then defined to be
\be
HF(M,k) = {\rm dim}(M_k) = {\rm dim}(R_k) -{\rm dim}(I_k) \quad.
\ee
The Hilbert series is then $H(M,t) = \sum_k HF (M,k)t^k$.

Usually, we would like to construct the quotient ring $M$ by finding a typical ideal, which is usually generated by a few polynomials
$f_1,...,f_s$. Then questions such as inclusion of a polynomial inside the ideal and non-trivial relations among generators, arise in this
process. The answers to these questions are computational and generally quite hard. However, a special set of basis of the ideal, called
\ep{Gr\"obner basis} can be constructed most efficiently to describe the polynomial sequence $f_1,...,f_s$. Now we denote the set
of polynomials in Gr\"obner basis by $g_1,...,g_r$, where $r \neq s$ in general. Since $f_i$ are taken as basis vectors for the ideal, we
can change the basis to the new Gr\"obner basis, which simply generate the same ideal. A more thorough treatment for this topic can be
found in \cite{Cox:1992ca}. To construct the Gr\"obner basis, we need to define an ordering of monomial first.\footnote{The common
	choices of monomial ordering are lexographic, graded lexographic and graded reverse lexographic ordering. Take two monomials
	$\textbf{x}^{\alpha_1}...x_n^{\alpha_n}$ and $\textbf{x}^{\beta_1}...x_n^{\beta_n}$ of total degree $\alpha=\alpha_1+\cdots+\alpha_n$ and
	$\beta=\beta_1+\cdots+\beta_n$. We take $\textbf{x}^\alpha > \textbf{x}^\beta$ if $\alpha >\beta$; if $\alpha = \beta$, then $\textbf{x}^\alpha > \textbf{x}^\beta$
	if $\alpha_1>\beta_1$; if $\alpha=\beta$ and $\alpha_1 = \beta_1$, then $\textbf{x}^\alpha > \textbf{x}^\beta$ if $\alpha_2 >\beta_2$ and so on.}

This monomial ordering ``$>$"
determines whether ${\bf x} ^\alpha > {\bf x} ^\beta, {\bf x} ^\alpha = {\bf x} ^\beta$ or ${\bf x} ^\alpha < {\bf x} ^\beta$ for two monomials 
${\bf x} ^\alpha = x_1^{\alpha_1}\cdots x_n^{\alpha_n}$ and ${\bf x} ^\beta= x_1^{\beta_1}\cdots x_n^{\beta_n}$. With this ordering, we can
then find the ``largest" monomial inside a polynomial $h \in K[x_1,...,x_n]$. This is defined to be the initial monomial of $h$, denoted by ${\rm in}(h)$.
\footnote{This is also commonly defined as the leading term of $h$ and denoted by LT($h$).} For every polynomial in $I = \braket{f_1,...,f_s}$,
we take their initial monomial and denote this set to be in($I$). In general, in($I$) is not equal to the set generated by initial monomials of the $f_i$.
But the defining property of Gr\"obner basis is that in($I$) = $\braket{\textrm{in}(g_1),...,\textrm{in}(g_r)}$.

With the above abstract definition, we shall benefit from a few concrete examples. First let us take the polynomial ring of two variables with
coefficients in the real numbers, $R = \mathbb{R}[x,y]$. Here we take monomial ordering to be graded reverse lexographic ordering, which
is the default setting for computer package \texttt{Macaulay2}.
\paragraph{Exmaple 1}
Let $R = \mathbb{R}]x,y]$ and $I =\braket{x+y}$. As the ideal has just a single polynomial, it is then by definition, a Gr\"obner basis. Hence,
the initial ideal is generated by the in$(x+y) = x$, in$(I) = \braket{x}$. The Hilbert series for the quotient ring $M = R/I$ is equivalent to the Hilbert
series of $R/\textrm{in}(I) = \mathbb{R}[x,y]/\braket{x} = \mathbb{R}[y]$. Therefore, we have
\[\ H(\mathbb{R}[x,y]/\braket{x+y}, t) = \frac{1}{1-t} \quad. \]\

\paragraph{Example 2}
Let $R=\mathbb{R}[x,y]$ and $I = <x^2,y^3>$. A monomial of the form $x^\alpha y^\beta$ is in the ideal for $\alpha \geq 2$ and $\beta \geq 3$.
Therefore, the monomials for the quotient ring are $1,x,y,xy,y^2,xy^2$. Since the Hilbert series counts the independent monomials, it is then
\[\ H(\mathbb{R}[x,y]/\braket{x^2,y^3}, t) = 1  + 2t + 2t^2 + t^3 \quad. \]\
This finite polynomial hints us that it can be written as a rational function with both numerator and denominator being in Euler form. This is actually
\[\
H = \frac{(1-t^2)(1-t^3)}{(1-t)^2} = \frac{1-t^2-t^3+t^5}{(1-t)^2} \quad \]\
where the denominator is the Hilbert series of free ring $\mathbb{R}[x,y]$, while the numerator reflects the relation among generators of the ideal.

\end{document}